\documentclass[11pt]{article}
\pdfoutput=1
\usepackage[margin=1.0in]{geometry}
\usepackage{amsmath,amssymb,graphicx}
\usepackage{hyperref}
\usepackage{slashed}

\usepackage[T1]{fontenc}

\usepackage[utf8]{inputenc}

\usepackage[greek,english]{babel}

\usepackage{framed}

\usepackage[font=small,labelfont=bf]{caption}
\usepackage{cite}

\newcommand{\iu}{{\mathrm i}}
\newcommand{\E}{{\mathrm e}}
\newcommand{\cpi}{\text{\greektext p}}

\newcommand{\RR}{{\mathbb{R}}}
\newcommand{\ZZ}{{\mathbb{Z}}}

\newcommand{\FM}{{F_\textsc{M}}}
\newcommand{\AM}{{A_\textsc{M}}}
\newcommand{\JE}{{J_\textsc{E}}}
\newcommand{\JM}{{J_\textsc{M}}}

\newcommand{\cF}{\mathcal{F}}
\newcommand{\cA}{\mathcal{A}}
\newcommand{\cFM}{\mathcal{F}_\textsc{M}}

\newcommand{\rmd}{\textrm{d}}
\newcommand{\dif}[1]{\ensuremath{\operatorname{d}\!{#1}}}

\newcommand{\Uone}{{\mathrm{U(1)}}}
\newcommand{\SU}{{\mathrm{SU}}}

\newcommand{\rhoM}{{\rho_\textsc{M}}}
\newcommand{\JvecM}{{{\bm J}_\textsc{M}}}
\newcommand{\gagg}{{g_{a\gamma\gamma}}}

\let\Re\undefined
\let\Im\undefined

\DeclareMathOperator{\Re}{Re}
\DeclareMathOperator{\Im}{Im}
\DeclareMathOperator{\Tr}{Tr}

\newcommand{\SL}{\mathrm{SL}}
\newcommand{\PSL}{\mathrm{PSL}}

\newcommand{\be}{\begin{equation}}
\newcommand{\ee}{\end{equation}}

\usepackage{xcolor}
\usepackage{bm}

\title{\bf  Non-standard axion electrodynamics\\ and the dual Witten effect}
\author{Ben Heidenreich${}^1$,  Jacob McNamara${}^2$, and Matthew Reece${}^3$ \\[10pt]
{\small ${}^1$Department of Physics, University of Massachusetts, Amherst, MA 01003 USA}\\
{\small ${}^2$Walter Burke Institute for Theoretical Physics}\\
{\small California Institute of Technology, Pasadena, CA 91125, USA}\\
{\small ${}^3$Department of Physics, Harvard University, Cambridge, MA, 02138, USA}
}

\date{November 29, 2023}

\begin{document}

\begingroup
{\flushright ACFI-T23-08\par}
\let\newpage\relax%
\maketitle
\endgroup

\begin{abstract}
Standard axion electrodynamics has two closely related features. First, the coupling of a massless axion field to photons is quantized, in units proportional to the electric gauge coupling squared. Second, the equations of motion tell us that a time-dependent axion field in a background magnetic field sources an effective electric current, but a time-dependent axion field in a background electric field has no effect. These properties, which manifestly violate electric-magnetic duality, play a crucial role in experimental searches for axions. Recently, electric-magnetic duality has been used to motivate the possible existence of non-standard axion couplings, which can both violate the usual quantization rule and exchange the roles of electric and magnetic fields in axion electrodynamics. We show that these non-standard couplings can be derived from SL(2,$\mathbb{Z}$) duality, but that they come at a substantial cost: in non-standard axion electrodynamics, all electrically charged particles become dyons when the axion traverses its field range, in a dual form of the standard Witten effect monodromy. This implies that there are dyons near the weak scale, leads to a large axion mass induced by Standard Model fermion loops, and dramatically alters Higgs physics. We conclude that non-standard axion electrodynamics, although interesting to consider in abstract quantum field theory, is not phenomenologically viable.
\end{abstract}

\tableofcontents

\section{Introduction and Central Argument}

In this paper, we study axion electrodynamics: the interaction of a periodic scalar field $\theta \cong \theta + 2\cpi$ (the axion) with a $\Uone$ gauge field $A$ (the photon) with field strength $F$ through a topological, Chern-Simons-type interaction:
\begin{equation} \label{eq:action}
\int \rmd^4x\, \sqrt{|g|}\left(-\frac{1}{2} f^2 \partial_\mu \theta \partial^\mu \theta - \frac{1}{4 e^2} F_{\mu \nu}F^{\mu \nu}\right) + \frac{n}{8\cpi^2} \int \theta F \wedge F,
\end{equation}
where in the last term we use the differential form notation $F = \frac{1}{2} F_{\mu\nu} \dif x^\mu \wedge \dif x^\nu$ to emphasize the topological nature of the interaction. It is a well-known fact that a consistent quantum field theory with this action obeys a quantization condition,
\begin{equation} \label{eq:quantization}
n \in \ZZ,
\end{equation}
where $A$ is normalized such that the minimally charged particle has charge 1, and we assume the spacetime background is restricted to spin 4-manifolds. This quantization condition has important applications for the couplings of axion fields of interest in real-world particle physics, including the QCD axion~\cite{Peccei:1977ur, Peccei:1977hh, Weinberg:1977ma, Wilczek:1977pj} or more general axion-like particles. Such particles are the subject of intense experimental scrutiny. Models where $n$ is an order-one integer provide natural targets of such experimental searches (though very large integer values of $n$ are also possible, in principle~\cite{Farina:2016tgd, Agrawal:2017cmd}). 

Given the action~\eqref{eq:action}, one can derive an axionic modification of Maxwell's equations~\cite{Sikivie:1983ip}, which takes the form:
\begin{equation}
\begin{aligned} \label{eq:axionMaxwell}
\bm{\nabla} \cdot \bm{E} &= \rho - \gagg \bm{B} \cdot \bm{\nabla} a\,,  &
\bm{\nabla} \times \bm{E} &= - \frac{\partial \bm{B}}{\partial t} - \JvecM \,, \\
\bm{\nabla} \cdot \bm{B} &= \rhoM \,, &
\bm{\nabla} \times \bm{B} &= \frac{\partial \bm{E}}{\partial t} + \bm{J} - \gagg \biggl(-\bm{B} \frac{\partial a}{\partial t} + \bm{E} \times \bm{\nabla}a \biggr) \,.
\end{aligned}
\end{equation}
Here $\rho, \bm{J}$ are the usual electric charge density and current, $\rhoM, \JvecM$ are the (hypothetical) magnetic charge density and current, $a(x) = f \theta(x)$ is the canonically normalized axion field, $\bm{E}, \bm{B}$ are the canonically normalized electric and magnetic fields, and 
\begin{equation} \label{eq:gagg}
\gagg = \frac{n e^2}{4\cpi^2 f}
\end{equation}
is the axion photon coupling, proportional to the integer $n$.\footnote{In a realistic model with additional interactions beyond those in~\eqref{eq:action}, there may be additional contributions of the form $n \mapsto n + \delta n$ where $\delta n \propto m_a^2$ is not quantized but vanishes in the limit of zero axion mass. Such contributions can be interpreted as $\Box \theta\, F \wedge F$ interactions. For the QCD axion, the contribution from the axion-pion mixing is an important example~\cite{Kaplan:1985dv, Srednicki:1985xd, Georgi:1986df, Svrcek:2006yi, Agrawal:2017cmd, Agrawal:2023sbp}. This is well-understood physics, distinct from our concerns in this paper.} 

The equations~\eqref{eq:axionMaxwell} manifestly break electric-magnetic duality. For example, a time-dependent axion field in a background magnetic field leads to an effective {\em electric} current, sourcing $\bm{\nabla} \times \bm{B}$. Many searches for axion dark matter rely on this coupling. Furthermore, we see that an axion gradient aligned with a magnetic field behaves as an effective {\em electric} charge density. The axion does not source effective magnetic charge densities or currents. This breaking of electric-magnetic duality is also reflected in the fact that it is the electric coupling $e$ that appears in the numerator of~\eqref{eq:gagg}, rather than the magnetic coupling (which is {\em inversely} proportional to $e$). The fact that axion electrodynamics breaks electric-magnetic duality has spurred some authors to propose non-standard formulations of axion electrodynamics, which aim to either restore electric-magnetic duality~\cite{Visinelli:2013mzg} or break it in alternative ways~\cite{Sokolov:2021ydn, Sokolov:2022fvs, Sokolov:2023pos}. These non-standard formulations of axion electrodynamics not only allow for $\gagg \propto 1/e^2$, implying much stronger couplings, they also introduce new terms in~\eqref{eq:axionMaxwell}; for example, allowing $\bm{E} \frac{\partial a}{\partial t}$ to source $\bm{\nabla} \times \bm{E}$. Some of the proposed modifications have begun to receive attention in the context of the design or interpretation of experiments~\cite{McAllister:2015zcz, Ko:2016izh, Tercas:2018gxv, Le:2021max, Bourhill:2022alm, Li:2022oel, Tobar:2022rko, McAllister:2022ibe, Li:2023kfh, Li:2023aow, Tobar:2023rga, Patkos:2023lof} or astrophysical observations~\cite{Anzuini:2022bqd}. Thus, it is important to understand to what extent such a non-standard axion electrodynamics is theoretically and phenomenologically viable. We will focus our attention on the formulation in~\cite{Sokolov:2022fvs}, which constitutes the bulk of this literature. (The alternative introduced in~\cite{Visinelli:2013mzg} is on a less sound footing since it does follow from any known action, but it shares the same new terms in the equations of motion that we will argue are phenomenologically excluded.) 

In this paper, we consider an in-principle well-motivated theoretical alternative to standard axion electrodynamics, namely, to implement a coupling of the form $\theta F' \wedge F'$ where $F'$ is an SL(2,$\ZZ$) dual of the standard field strength $F$.\footnote{To be precise, the coupling takes this form in the $F'$ duality frame. See~\S\ref{subsec:alternative} for the form of the coupling in the $F$ duality frame.} We show that such a coupling leads to equations of motion that can be written in the form studied in~\cite{Sokolov:2022fvs}. However, these equations have an important implication. In non-standard axion electrodynamics, every electrically charged particle undergoes a monodromy, becoming a dyon in the presence of an axion field that evolves around the circle from $\theta = 0$ to $\theta = 2\cpi$. We will argue that this is inconsistent with the physics of our universe, and in particular with the existence of light, weakly coupled, chiral fermion fields that obtain a mass only from electroweak symmetry breaking. Thus, although non-standard axion electrodynamics is interesting from the viewpoint of quantum field theory, it is already excluded as a theory of real-world particle phenomenology. Throughout the paper, we use standard quantum field theory formalism, rather than the less standard Zwanziger approach that appears in recent work like~\cite{Sokolov:2022fvs}. Nothing is lost by doing so, but to reassure devotees of that formalism, we emphasize that our key results rely only on the equations of motion away from singular sources, not on the precise fashion in which these (massive) sources are quantized.


\begin{figure}[!h]
\centering
\includegraphics [width = 0.5\textwidth]{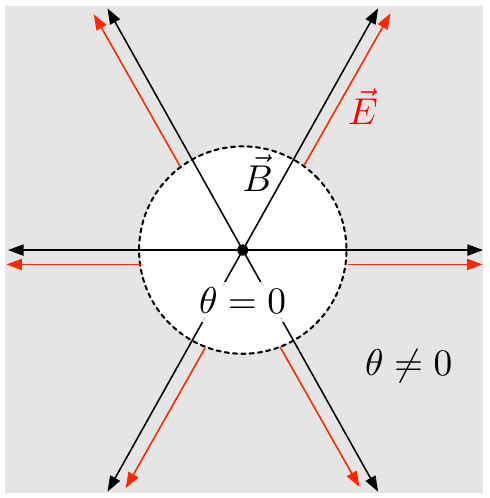}
\caption{A classic argument for the Witten effect~\cite{Wilczek:1987mv}. A magnetic monopole in a region of zero $\theta$ appears to be a dyon far away in a region with $\theta \neq 0$.
} \label{fig:witteneffect}
\end{figure}

\begin{figure}[!h]
\centering
\includegraphics [width = \textwidth]{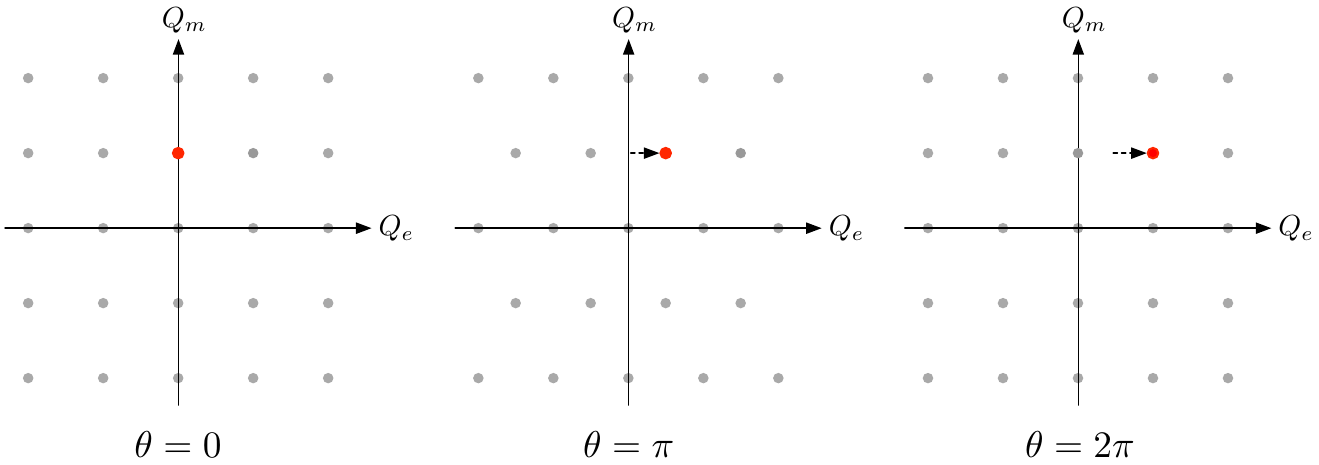}
\caption{The Witten monodromy. As $\theta$ increases the monopole (red point) gradually acquires electric charge, ending up with a full charge quantum at $\theta = 2\pi$. The complete dyon spectrum (gray points) has now returned to its original configuration, reflecting the periodicity of $\theta$, even though individual dyons have different charges than they started out with.} \label{fig:wittenmonodromy}
\end{figure}

Let us now sketch out our argument, to be explained more precisely in subsequent sections. Our reasoning relies crucially on the Witten effect~\cite{Witten:1979ey}: in an environment with nonzero $\theta$, a magnetic monopole with unit magnetic charge acquires a fractional electric charge $\frac{n\theta}{2\cpi}$. A simple argument for this (originating in~\cite{Wilczek:1987mv}; also see~\cite{TongGT}), is to consider a monopole carrying purely magnetic charge in a local environment with zero $\theta$, surrounded by a region in which a nonzero value of $\theta$ turns on at larger radius. The equation for $\bm{\nabla} \cdot \bm{E}$ implies that the radial $\bm{B}$ field sourced by the monopole, together with the radial axion gradient $\bm{\nabla} \theta$, will source an electric field at larger radii. (See Fig.~\ref{fig:witteneffect}.) Thus, an observer at larger distances will see an electric field that appears to have been sourced by a particle with nonzero electric charge. If we shrink the size of the region around the monopole with $\theta = 0$, the effective charge observed from afar doesn't change. Thus, in the limit that we embed the monopole in an environment with constant $\theta$ everywhere we conclude that it is a dyon, of electric charge $\frac{n\theta}{2\cpi}$. If we continuously vary $\theta$ from $0$ to $2\cpi q$ (with $q \in \ZZ$), a magnetic monopole becomes a dyon with $nq$ full units of electric charge, even though (due to its periodicity) the $\theta$ value in its environment has returned to its starting point. In general, the dyon's mass will increase in this process. This phenomenon, in which the theory is periodic as a function of $\theta$ but a given particle will transmute into other particles when $\theta$ continuously varies around its circle, is known as ``monodromy,'' and it arises in contexts as simple as the familiar problem of a quantum-mechanical particle on a circle (reviewed in, e.g.,~\cite{Gaiotto:2017yup,Cordova:2019jnf}). Other straightforward arguments for the Witten effect, independent of the UV completion of the theory, appear in~\cite{Coleman:1982cx, Preskill:1984gd}. Because it plays a central role in our argument, below we will use the phrase ``Witten monodromy'' to mean the monodromy in the dyon spectrum under $\theta \to \theta + 2\cpi$ induced by the Witten effect. (See Fig.~\ref{fig:wittenmonodromy}). To dispel any lingering doubts, in~\S\ref{subsec:witten} we will review an argument that does not refer to magnetic monopoles at all but solely focuses on the construction of a dual magnetic gauge field in regions away from point charges, which directly demonstrates the Witten monodromy. 

Now, suppose that rather than the standard equations~\eqref{eq:axionMaxwell}, we had a modified equation in which $\bm{\nabla} \cdot \bm{B}$ is sourced by a term of the form $\bm{E} \cdot \bm{\nabla} a$. Such a term appears explicitly in the proposed modified equations in~\cite{Visinelli:2013mzg, Sokolov:2022fvs}. It leads to a magnetic dual of the Witten effect: in a $\theta$ background, such a term would imply that a particle that has purely {\em electric} charge in a region of zero $\theta$ acquires an effective {\em magnetic} charge $\frac{n\theta}{2\cpi}$ when in a region of nonzero $\theta$. One might reasonably ask if this is a well-defined claim. It is perfectly reasonable, and even standard, to {\em define} electric charge to be the charge carried by an electron, so that it carries zero magnetic charge by definition. Indeed, it is well-known that there are several different useful ways to define charge in the presence of Chern-Simons terms~\cite{Marolf:2000cb}. However, independent of one's preferred definitions, an invariant physical fact remains: there is a dual Witten monodromy. That is to say, if we continuously vary $\theta$ from $0$ to $2\cpi q$ (with $q \in \ZZ$), the ordinary electron would become a dyon state with magnetic charge $nq$. Because these are two different states in the same theory, this is an invariant physical fact, not an artifact of a particular definition of charge.

Unlike the standard Witten effect, this dual monodromy is a phenomenological disaster. We claim that it is completely impossible. In the world around us, we do not observe a collection of light dyon states with the mass of the electron and arbitrary amounts of magnetic charge. Thus, in the process of varying $\theta$ from $0$ to $2\cpi$ and turning the electron into a dyon, the electron mass should increase (dramatically!) as $\theta$ turns on. As a result, electron loops would generate a large perturbative mass for the axion.\footnote{If the axion in question were the QCD axion, this effect would dominate over the contribution from QCD instantons and spoil the solution to the Strong CP problem.} However, this is only the start of the problem, as the electron is a {\em chiral} fermion in the Standard Model. The electron obtains a mass only via electroweak symmetry breaking, and $\theta$ is a neutral scalar, so turning it on cannot violate electroweak symmetry. At best, we can couple $\theta$ to a Higgs-dependent electron mass term. This implies an infinite tower of dyon states all obtaining a mass from the Higgs, which would drive the Higgs field to strong coupling and significantly alter Standard Model predictions for Higgs properties.\footnote{We expect that such a theory, with an infinite tower of states obtaining mass from the Higgsing of a nonabelian gauge theory, is actually inconsistent even at the formal level. However, even if such a theory exists formally, it is certainly not compatible with observed physics.} All said, there is no way to modify axion electrodynamics and obtain anything resembling the Standard Model coupled to a light axion.

This is our central argument: modifying axion electrodynamics would require that the electron (and every other elementary charged particle) obtains a magnetic charge in an axion background, which is impossible due to the chiral structure of the Standard Model and the desire for a light axion. Before returning to this point, we will first review the physics of axion electrodynamics and electric-magnetic duality in more detail below, in the interest of providing a clear pedagogical reference and a more complete argument. We will highlight some other interesting and under-appreciated physics along the way. The outline of the paper is as follows: in~\S\ref{sec:standard}, we review standard axion electrodynamics and prove the quantization condition~\eqref{eq:quantization}. We also give a straightforward derivation of the Witten monodromy.
In~\S\ref{sec:magnetic}, we discuss electric-magnetic duality and explain how it allows non-standard axion electrodynamics evading the quantization condition in the context of $\Uone$ gauge theory with no charged matter coupled to an axion. In~\S\ref{sec:phenoassess}, we argue that non-standard axion electrodynamics is incompatible with the Standard Model (for the reason we have just explained above). Finally, in~\S\ref{sec:conclusions} we offer some concluding remarks. In appendix~\ref{app:backwards} we systematically compare our approach with \cite{Sokolov:2022fvs} and derive quantization rules for the generalized axion couplings.

\section{Standard Axion Electrodynamics}
\label{sec:standard}

In this section we review standard axion electrodynamics and derive its various features, such as the Witten monodromy. Readers interested in a more in-depth treatment of many of these ideas may also wish to consult the TASI lectures~\cite{Reece:2023czb} by one of the authors. 

\subsection{Derivation of coupling quantization}
\label{subsec:derivation}

We begin by giving the simple derivation that the action~\eqref{eq:action} only defines a consistent quantum field theory when $n \in \ZZ$. A consistent quantum field theory can be studied on a variety of spacetime backgrounds. This is a necessity for theories that can be consistently coupled to gravity. In particular, we will consider the Euclidean continuation of the theory on a 4-manifold (without boundary) $M$, in which the Chern-Simons term $\int \theta F \wedge F$ acquires an extra factor of $\iu$. Our quantum field theory is defined by a path integral summing over field configurations for $\theta$ and $A$. We will begin with four key assumptions:
\begin{enumerate}
\item The axion field is periodic (we often say it is a ``compact scalar''): $\theta \cong \theta + 2\cpi$. In particular, this allows for field configurations in which the value of $\theta$ {\em winds} around a circle in spacetime. This means that $\theta$ itself is not a well-defined (gauge invariant) variable, whereas $\E^{\iu \theta}$ is. We can think of $\theta \mapsto \theta + 2\cpi$ as a gauge transformation.
\item The photon's gauge group is $\Uone$, which is compact. Gauge transformations take the form $A \mapsto A + \iu g^{-1} \dif{g}$, where $g(x) = \E^{\iu \alpha(x)}$ takes values in $\Uone$. The distinction between this and the related non-compact gauge group $\RR$, both of which have the Lie algebra $\mathfrak{u}(1) \cong \RR$, is that the gauge transformations for $\Uone$ can wind around circles in spacetime, allowing for non-trivial disorder operators such as 't~Hooft lines. 
\item The axion field $\theta$ is invariant under $\Uone$ gauge transformations of $A$.
\item The gauge field $A$ (along with its field strength $F$) is invariant under the $2\cpi$ shift of $\theta$.
\end{enumerate}
The path integral sums over all field configurations for $\theta$ and $A$, which, because of their respective periodicity properties, include topologically nontrivial field configurations. For example, field configurations can have a winding number of the axion around a 1-cycle $C$:
\begin{equation} \label{eq:winding}
\frac{1}{2\cpi} \int_C \dif\theta = w(C) \in \ZZ
\end{equation}
and a magnetic flux of the gauge field through a 2-cycle $S$:
\begin{equation} \label{eq:flux}
\frac{1}{2\cpi} \int_S F = m(S) \in \ZZ.
\end{equation}
In more mathematical jargon, we can think of $w(C)$ and $m(S)$ as information about classes in integer cohomology, $[\frac{1}{2\cpi} \dif \theta] \in H^1(M, \ZZ)$ and $[\frac{1}{2\cpi} F] \in H^2(M, \ZZ)$.\footnote{More precisely, the quantized flux $m(S) \in \mathbb{Z}$ determine the free part of the integral cohomology class $[\frac{1}{2\cpi} F] \in H^2(M, \ZZ)$, while the torsion part is encoded in the holonomies of $A$. There is no analogous subtlety for $w(C)$, since $H^1(M,\mathbb{Z})$ is torsion-free for \emph{any} topological space $M$. This subtlety has no effect on our subsequent discussion.} As in the familiar case of the Dirac monopole, a nontrivial topology means that we can't define the fields $\theta$ and $A$ globally, but we can patch them together on different coordinate charts such that, on overlaps, they agree up to gauge transformations. The field strengths $\dif \theta$ and $F$ are defined globally. Once we specify any field configuration $(\theta, A)$ lying in a particular cohomology class, then the {\em differences} $(\theta - \theta', A - A')$ between this and any other field configuration $(\theta', A')$ specified by the same classes are globally well-defined. This allows us to separate the path integral into a discrete sum over topological classes, together with a continuous integral over field configurations without regard to topology.

Now we rely on a mathematical fact that we will not prove (see, e.g.,~\cite{BottTu, Nakahara:2003nw}): if a 2-form $\omega$ is a representative of a class in integer cohomology, then the 4-form $\omega \wedge \omega$ is {\em also} a representative of a class in integer cohomology. In other words, once we have chosen an $F$ such that~\eqref{eq:flux} holds, we are also guaranteed that
\begin{equation} \label{eq:FF1}
\frac{1}{4\cpi^2} \int_M F \wedge F \in \ZZ \qquad \text{(any $M$)}.
\end{equation}
This is sufficient to derive a quantization condition on axion-photon couplings, but we can do slightly better. For describing real-world physics, we can restrict to spacetime manifolds on which it is possible to define fermion fields. These are known as spin manifolds, and it turns out that on a spin manifold the integer~\eqref{eq:FF1} is always even. That is, we have:
\begin{equation} \label{eq:FF}
\frac{1}{8\cpi^2} \int_M F \wedge F \in \ZZ \qquad \text{(any {\em spin} $M$)}.
\end{equation}
Now, the action~\eqref{eq:action} is manifestly invariant under $\Uone$ gauge transformations but is not invariant under the shift $\theta \mapsto \theta + 2\cpi$. However, physical quantities depend only on the exponentiated Euclidean action, $\exp(-S_E[A, \theta])$, because this appears in the path integral measure. We have:
\begin{equation} \label{eq:invariancecheck}
\theta \mapsto \theta + 2\cpi: \qquad \E^{-S_E[A, \theta]} \mapsto \E^{-S_E[A, \theta]}  \exp\left[{-\frac{\iu n}{4 \cpi} \int F \wedge F}\right].
\end{equation}
Now, for {\em every} field configuration that we sum over in the path integral, the integral appearing in the last factor of~\eqref{eq:invariancecheck} is of the form $8\cpi^2 k$ for some $k \in \ZZ$, and hence the factor takes the form $\exp[-2\cpi \iu n k]$. This is always $1$ if $n \in \ZZ$, but in general is not $1$ if $n \notin \ZZ$. This proves~\eqref{eq:quantization}.

\subsection{Revisiting the assumptions}
\label{subsec:revisiting}

Our proof was straightforward, but relied on four assumptions. Let's revisit them one by one:
\begin{enumerate}
\item The axion was assumed to be periodic. If $\theta$ is a non-compact field, there is no $\theta \mapsto \theta + 2\cpi$ gauge redundancy, and the whole argument falls apart. On the other hand, we have good reasons for studying periodic axion fields, beyond the fact that compactness is often taken to be part of the definition of an axion. UV completions give rise to compact axions: a pseudo-Nambu-Goldstone boson of an approximate $\Uone$ global symmetry, or a zero mode of a higher-dimensional $\Uone$ gauge field, is intrinsically compact. A compact scalar can only admit periodic terms in its potential,\footnote{An exception is when the potential has different branches interchanged by monodromy, so that $\theta$ is effectively non-compact~\cite{Silverstein:2008sg, Kaloper:2008fb}. This is still highly constrained, because it arises from a $\theta F_4$ coupling with a quantized coefficient. For a QCD axion, this coefficient is expected to be zero. Otherwise, the resulting potential will dominate over the QCD contribution to the axion potential, but will generically have a minimum in a different location (spoiling the solution to the Strong CP problem).} which opens up the possibility that the potential is dominated by exponentially small instanton effects.\footnote{This is only a possibility, not a guarantee. The axion quality problem is essentially the question of why additional periodic terms in $V(\theta)$ with large coefficients do not exist. This problem has at least one highly effective solution, which is to posit that the axion is a zero mode of a higher-dimensional gauge field~\cite{Witten:1984dg, Choi:2003wr}. The situation is much worse for non-compact scalars. For example, in a supersymmetric theory, the non-compact saxion will generically obtain a mass from K\"ahler potential terms in the presence of SUSY breaking, while the axion can remain exponentially lighter (see, e.g.,~\cite{Conlon:2006tq}).} For a generic non-compact scalar, it would be difficult to explain why the field is light and why its dominant source of shift-symmetry breaking originates from coupling to gluons, so it is unlikely to solve the Strong CP problem. In short, if we want to drop the compactness assumption on the axion, we are not considering a traditional axion at all, and we have to modify the entire structure of the model.
\item The gauge group was taken to be $\Uone$ rather than $\RR$. If this assumption is dropped, there can be no magnetic flux, $\int F = 0$ for any 2-cycle, and $\int F \wedge F = 0$ for any 4-manifold. Then the axion-photon coupling can take on any real value. However, there are compelling arguments that consistency of black hole physics forbids $\RR$ gauge groups from arising in quantum gravity~\cite{Banks:2010zn}, so we do not expect this case to be relevant in the real world.\footnote{It should also be noted that the non-standard axion couplings considered in~\S\ref{sec:magnetic} are impossible if the electromagnetic gauge group is $\mathbb{R}$, because only U$(1)$ has the necessary SL$(2,\mathbb{Z})$ self-duality required to make the non-standard axion periodic.}
\item The axion field $\theta$ was assumed to be invariant under $\Uone$ gauge transformations. If it were not, it would get eaten via the Stueckelberg mechanism, and give the photon a mass. To make $\exp(\iu S)$ gauge invariant, we would have to add anomalous charged matter as in the 4d Green-Schwarz mechanism. A massive photon scenario is not the case of interest for us, but because our argument made use of invariance under $\theta$ gauge transformations but not $A$ gauge transformations, dropping this assumption would also not change the conclusion.
\item The gauge field strength $F$ was assumed not to change under the gauge transformation $\theta \mapsto \theta + 2\cpi$. This may seem innocuous, but in fact it is the weakest point in the argument. We will explain the possible alternative, a dual Witten monodromy, in \S\ref{sec:magnetic}.
\end{enumerate}

The first two assumptions can't be evaded by flowing from a UV theory in which they hold to an IR theory in which they do not. If one begins with multiple $\Uone$ gauge fields and higgses, the surviving massless gauge field has a compact $\Uone$ gauge group. Similarly, if one begins with multiple axion fields and then gives a mass to some of them, either via a periodic potential or through a Stueckelberg mechanism in which they are eaten by a gauge field, a surviving light axion is always compact~\cite{Choi:2019ahy, Fraser:2019ojt}. This fact has proven useful in diagnosing some mistaken analyses of multi-axion models in the literature.

\subsection{The Witten monodromy and anomaly inflow}
\label{subsec:witten}

Next, we show that the monodromy associated with the Witten effect for a dynamical axion can be derived in a very straightforward way, without referring to pointlike monopoles at all. This makes it clear that it is an effect within the low-energy effective field theory associated with the action~\eqref{eq:action}, independent of details of the UV completion (in contrast to some claims~\cite{Sokolov:2022fvs}). This argument is simply a special case of the much more general phenomenon of {\em anomaly inflow} in the presence of Chern-Simons terms~\cite{Callan:1984sa}; this was also recently pointed out in~\cite{Fukuda:2020imw}. 

To derive the Witten monodromy, let's first recall what it means to introduce a magnetic dual gauge field $\AM$. The field strength of the magnetic dual gauge field should be the Hodge dual of the usual gauge field strength, up to normalization. Specifically, in free Maxwell theory without a $\theta$ term and without an axion coupling, we would define
\begin{equation} \label{eq:introduceAM}
\frac{1}{2\cpi} \dif\AM = -\frac{1}{e^2} \star F.\qquad\text{(no axion)}
\end{equation}
The integral of the left hand side gives the magnetic flux of the gauge field $\AM$, which is minus the electric flux of the original gauge field $A$, which we know to be measured by the right-hand side. Now, the reason that the equation~\eqref{eq:introduceAM} makes sense is that Maxwell's equations, in the absence of any electric charges or currents, tell us that $\dif \star F = 0$, i.e., that the electric flux density $\star F$ is {\em closed}. Any closed form is {\em locally} exact, which means that in any given region, we can find a solution $\AM$ to the equation~\eqref{eq:introduceAM}. There is no guarantee that $\star F$ is exact, which means that we may not be able to {\em globally} define $\AM$, but this is fine: we can define it locally in different coordinate patches, with agreement on the overlaps to construct a gauge bundle. Also, solutions to~\eqref{eq:introduceAM} are not unique: if $\AM$ solves the equation, so does $\AM - \dif\alpha_\textsc{M}$ for any $\alpha_\textsc{M}$. This is the expected magnetic gauge redundancy. Gauge transformations of $A$ do not act on $\AM$, and vice versa.

For axion electrodynamics with the action~\eqref{eq:action} (and $n = 1$, for simplicity), introducing $\AM$ is not so straightforward. The reason is that we now have an equation of motion 
\begin{equation} \label{eq:axioneom}
\frac{1}{e^2} \dif{\star F} = \frac{1}{4\cpi^2} \dif \theta \wedge F.
\end{equation}
The electric flux density $\star F$ is no longer closed, even away from charged particles, in the presence of a varying axion field. This means that we can no longer find a solution to~\eqref{eq:introduceAM}; it is simply not the right way to locally define a magnetic gauge field $\AM$. However, we can rewrite~\eqref{eq:axioneom} in the form
\begin{equation} \label{eq:closedaxioncase}
\dif{\left(\frac{1}{e^2} \star F - \frac{1}{4\cpi^2} \theta F\right)} = 0,
\end{equation}
which is equivalent away from magnetic monopoles where $\dif F \neq 0$. We have only been discussing equations that hold locally away from charged objects, so this restriction is fine. It motivates introducing the magnetic gauge field $\AM$ with the new definition
\begin{equation} \label{eq:introduceAMaxion}
\frac{1}{2\cpi} \dif\AM = -\frac{1}{e^2} \star F + \frac{1}{4\cpi^2} \theta F.\qquad\text{(with axion)}
\end{equation}
Just as before, we can always {\em locally} solve this equation, thanks to~\eqref{eq:closedaxioncase}. Again, solutions are not unique, which corresponds to the gauge redundancy of $\AM$. Furthermore, gauge transformations of $A$ do not affect $\AM$. However, we now have a new subtlety: the equation that we are solving for $\AM$ depends on $\theta$, which is itself not gauge invariant. In particular, if we construct a solution $\AM^{(0)}$ to~\eqref{eq:introduceAMaxion} and then perform a gauge transformation $\theta \mapsto \theta + 2\cpi$, our original $\AM^{(0)}$ will no longer be a solution. Instead, we have a new solution $\AM^{(1)} = \AM^{(0)} + A$. Said differently, the magnetic gauge field $\AM$ is {\em not gauge invariant} under the gauge transformation $\theta \mapsto \theta + 2\cpi$. It transforms as:
\begin{equation} \label{eq:wittenmonodromy}
\theta \mapsto \theta + 2\cpi:  \qquad \AM \mapsto \AM + A.
\end{equation} 
This result is the key equation specifying the Witten monodromy: an object with pure magnetic charge acquires one unit of electric charge under a complete shift of the axion around its field space.

This derivation of the Witten monodromy~\eqref{eq:wittenmonodromy} is very clean, since we only asked about how to define a magnetic gauge field {\em away} from any sources like monopoles or electrons. Thus, it is clear that the result has nothing to do with any divergences one might find in the cores of such objects, or any limiting procedure as in the argument we reviewed in the introduction. Nonetheless, it also implies the standard claims about dyonic modes on a magnetic monopole, through an anomaly inflow argument. A heavy magnetically charged object can be described by an effective theory living on its worldline $C$. Ordinarily, the dependence of the action of this object on $\AM$ would look like $S_\textsc{M} = \int_C \AM$. This is not a gauge-invariant action, but it is invariant when exponentiated, just as a standard Wilson loop is. However, in axion electrodynamics this is no longer true, because $\exp(\iu S_\textsc{M})$ is not invariant under~\eqref{eq:wittenmonodromy}. To fix this, we must add additional ingredients to our theory that cancel out the change in $S_\textsc{M}$. A minimal approach is to add a compact boson $\sigma \cong \sigma + 2\cpi$ that shifts under an $A$ gauge transformation, i.e.,
\begin{equation} \label{eq:dyonicmodeshift}
A \mapsto A - \dif \alpha: \qquad \sigma \mapsto \sigma - \alpha.
\end{equation}
This allows us to define a consistent worldline action
\begin{equation} \label{eq:SManomalyinflow}
S_\textsc{M} = \int_C \left[\AM - \frac{\theta}{2\cpi} (\dif \sigma + A)\right].
\end{equation}
(The full action $S_\textsc{M}$ will also include a monopole mass term that depends on the proper length of $C$ as well as a kinetic term for $\sigma$, but these are not relevant for our current discussion, which focuses only on charges.) By construction,~\eqref{eq:SManomalyinflow} is invariant under both $A$ gauge transformations and $\theta$ gauge transformations. The degree of freedom $\sigma$ behaves as a quantum-mechanical particle on a ring, which is the familiar dyonic degree of freedom on the monopole (originally discovered in the context of the 't~Hooft-Polyakov monopole~\cite{Jackiw:1975ep}). Here we see that the existence of this degree of freedom, or some other one with a similar ability to cancel the change in $S_\textsc{M}$ under $\theta \mapsto \theta + 2\cpi$, is a fundamental consistency requirement on the theory.\footnote{One might wonder if this argument can be evaded by imposing a $\theta = 0$ boundary condition on the monopole worldline. However, for dynamical monopoles (as opposed to 't~Hooft lines), this implies a strong coupling of the monopole to the axion. In fact, it is not really an alternative theory at all, it is just the limiting case where dyonic excitations become infinitely heavy (equivalently, the $\sigma$ kinetic term goes to zero). This is not an innocuous limit to take. For instance, monopole loop effects on the axion (as in~\cite{Fan:2021ntg}) are not exponentially suppressed in this limit.}

This argument is a particular case of a very general phenomenon: dualizing gauge fields in the presence of Chern-Simons terms produces magnetic gauge fields that are not invariant under electric gauge transformations. Consistency then requires that magnetically charged objects admit zero modes that can be excited to give them electric charge. These modes are said to arise by anomaly inflow~\cite{Callan:1984sa}. An exactly analogous argument tells us that axion strings admit chiral charge-carrying excitations.\footnote{Anomaly inflow arguments are often phrased in terms of a cancellation between a bulk anomaly and a localized anomaly, whereas we have phrased our argument in terms of a cancellation on the worldline. These pictures are equivalent, since the shift of $\AM$ arises from the bulk Chern-Simons term. Similarly, the bulk anomaly in the classic Callan-Harvey example of anomaly inflow on axion strings can be rephrased in terms of additional gauge transformations of the $B$ field on the string worldsheet; see, e.g., appendix B of~\cite{Heidenreich:2021yda} for a recent discussion.} An even more well-known example, with ample experimental verification, is the existence of edge modes in quantum Hall systems, which are described by Chern-Simons terms in $(2+1)$d with chiral electrically charged modes on the $(1+1)$d boundary.

The monodromy~\eqref{eq:wittenmonodromy}---rather than the details of the localized worldline mode required by anomaly inflow---will play the key role in our arguments below. Before moving on, let us make two other brief comments about the Witten effect. First, one might wonder what would have happened if we had traded the $\theta F$ term in~\eqref{eq:closedaxioncase} for an $A \wedge \dif\theta$ term. In this case, $\AM$ would have been defined differently, and would directly shift under an ordinary electric gauge transformation. One can work through the details, and find that (despite different intermediate steps) the physical conclusions are the same. Second, our argument above was about axion electrodynamics, and in particular $\dif \theta$ played a key role in the discussion starting from~\eqref{eq:axioneom}. The Witten effect in a theory with a constant $\theta$ term, rather than a dynamical axion, is slightly more subtle. Nonetheless, it can again be derived from general principles. Perhaps the most straightforward way to convince oneself of its validity is to dimensionally reduce to 2d QED with a $\theta$ term by compactifying on a closed 2-manifold with flux, then study the 2d theory on a spatial circle. This theory is equivalent to the quantum mechanics of a particle on a ring with a $\theta$ term, which is a familiar (and straightforward) problem to solve. The Witten effect here appears in the fact that the canonical momentum shifts in the presence of a nonzero $\theta$. As a result, the entire spectrum of the quantum mechanical theory is $\theta$ dependent, and exhibits monodromy. In fact, this textbook problem in ordinary quantum mechanics is exactly the same as the theory on the monopole worldline.    

\section{Duality and Non-Standard Axion Electrodynamics}
\label{sec:magnetic}

In this section, we review electric-magnetic duality in order to motivate non-standard axion-photon couplings which evade the formal arguments presented in \S\ref{subsec:derivation}. In particular, we find the possibility of greatly enhanced axion-photon couplings proportional to $1/e^2$, in agreement with the results of~\cite{Sokolov:2022fvs}. However, we also find that precisely when these enhanced axion-photon couplings appear, a dual version of the Witten monodromy leads to electric charges acquiring a magnetic charge when we take $\theta \mapsto \theta + 2\cpi$.

\subsection{Electric-magnetic duality for a free photon}

It is well-known that the theory of a free $\Uone$ gauge field has an SL(2,$\ZZ$) duality group, generated by the matrices
\begin{equation}
    S = \begin{pmatrix} 0 & 1 \\ -1 & 0 \end{pmatrix}, \quad T = \begin{pmatrix} 1 & 1 \\ 0 & 1 \end{pmatrix}.
\end{equation}
A general element of SL(2,$\ZZ$) has the form
\begin{equation} \label{eq:sl2z}
\Lambda = \begin{pmatrix} a & b \\ c & d \end{pmatrix}, \quad a, b, c, d \in \ZZ, \quad ad - bc = 1.
\end{equation}
The electric and magnetic potentials transform in a 2-dimensional representation:\footnote{To be clear, we are not considering a formulation of the theory where $A$ and $\AM$ are both integrated over in the path integral. One should really think of this equation as a shorthand for the transformation of physical quantities like Wilson and 't~Hooft lines, and electric and magnetic fluxes.} 
\begin{equation} \label{eq:ASL2transf}
\begin{pmatrix} \AM' \\ A' \end{pmatrix} = \begin{pmatrix} a & b \\ c & d \end{pmatrix} \begin{pmatrix} \AM \\ A \end{pmatrix} ,
\end{equation}
whereas the electric and magnetic currents $\JE$ and $\JM$ transform in the dual representation:
\begin{equation} \label{eq:JSL2transf}
\begin{pmatrix} \JM' & \JE' \end{pmatrix} = \begin{pmatrix} \JM & \JE \end{pmatrix} \Lambda^{-1} =  \begin{pmatrix} \JM & \JE \end{pmatrix} \begin{pmatrix} d & -b \\ -c & a \end{pmatrix},
\end{equation}
ensuring that the coupling $A \wedge \JE + \AM \wedge \JM$ is SL(2,$\ZZ$) invariant. Another equivalent way to write~\eqref{eq:JSL2transf} is
\begin{equation}
\begin{pmatrix} \JE' \\ -\JM' \end{pmatrix} = \begin{pmatrix} a & b \\ c & d \end{pmatrix}  \begin{pmatrix} \JE \\ -\JM \end{pmatrix}.
\end{equation}
The coupling constant and $\theta$ angle are packaged into a complex background field
\begin{equation} \label{eq:tau}
\tau = \frac{\theta}{2\cpi} + \iu \frac{2\cpi}{e^2}.
\end{equation}
This transforms as
\begin{equation}
   S: \tau \mapsto -\frac{1}{\tau}, \quad T: \tau \mapsto \tau + 1,
\end{equation}
or more generally 
\begin{equation} \label{eq:tautransf}
\tau \mapsto \frac{a\tau + b}{c\tau+d}
\end{equation}
under the matrix~\eqref{eq:sl2z}. The $T$ operation corresponds to a $2\cpi$ shift of $\theta$.

In the duality frame where our fundamental gauge field is $A$, we have electric and magnetic field strengths
\begin{equation} \label{eq:FandFM}
    F = \dif A, \quad \FM = -\frac{2\cpi}{e^2} \star F + \frac{\theta}{2\cpi} F,
\end{equation}
where $\FM = \dif \AM$ (where the gauge field $\AM$, like $A$ itself, need only be locally well-defined, i.e., it is a connection on a $\Uone$ bundle rather than a 1-form globally). Here we see explicitly that under the $T$ operation $\theta \mapsto \theta + 2\cpi$, we have $\FM \mapsto \FM + F$ and hence $\AM \mapsto \AM + A$. This is exactly the Witten monodromy~\eqref{eq:wittenmonodromy} that we derived in \S\ref{subsec:witten}, which we see is intrinsically part of the standard SL(2,$\ZZ$) formulation of electromagnetic duality. The magnetic flux quantization condition~\eqref{eq:flux} holds as a topological constraint on the $A$ field configurations we sum over in the path integral, whereas the analogous electric flux quantization condition
\begin{equation} \label{eq:eflux}
   -\frac{1}{2\cpi} \int_S \FM = \int_S \left(\frac{1}{e^2} \star F - \frac{\theta}{4\cpi^2} F\right) = e(S) \in \ZZ
\end{equation}
holds by the equations of motion. The appearance of $\theta$ in this condition, which is a direct consequence of the equations of motion, is one manifestation of the Witten effect.

It is possible to derive the SL(2,$\ZZ$) invariance directly from the path integral. The $T$ operation is the $2\cpi$ shift of $\theta$, which leaves the theory invariant for the reason we derived in the previous section. The $S$ operation can be derived by integrating in additional fields in the path integral and then integrating out all but one of the new fields~\cite{Witten:1995gf, Deligne:1999qp}. The partition function is not SL(2,$\ZZ$) invariant, but rather transforms as a modular form~\cite{Witten:1995gf}. After this operation one finds that the electric flux quantization condition~\eqref{eq:eflux} is now a topological constraint on the dual gauge field configurations that we now sum over in the path integral, whereas the magnetic flux quantization condition~\eqref{eq:flux} now holds via equations of motion. Thus, the question of whether flux quantization is topological or dynamical is not an invariant fact in $\Uone$ gauge theory, but an artifact of a chosen duality frame.

The theory of a free $\Uone$ gauge field has no particles with electric or magnetic charge, but it does have line operators which one can think of as infinitely heavy, static electrically and magnetically charged objects with which one can probe the theory. In particular, there is a Wilson line operator defined for integer $q$ (corresponding to a representation of $\Uone$) and curves $C$:
\begin{equation} \label{eq:Wilson}
    W_q(C) = \exp\left[\iu q \int_C A \right]
\end{equation}
and an 't~Hooft line operator $T_p(C)$ defined for a {\em magnetic} charge $p \in \ZZ$ and curve $C$. In terms of the magnetic dual gauge field, the 't~Hooft line operator is similar to a Wilson operator: 
\begin{equation} \label{eq:tHooft}
    T_p(C) = \exp\left[\iu p \int_C \AM\right].
\end{equation}
In terms of the electric gauge field $A$ over which we perform the path integral in the standard duality frame, we can define the 't~Hooft operator by excising a small tube around $C$ and imposing a boundary condition on the field that the magnetic flux $m(S)$ through a surface $S$ linking $C$ with linking number $\ell(S,C)$ is $\ell(S,C) p$. One further has an infinite collection of dyonic line operators $L_{p,q}(C)$ with magnetic charge $p$ and electric charge $q$, which can be thought of as the fusion of $q$ minimal-charge Wilson lines and $p$ minimal-charge 't~Hooft lines. The duality group acts on this collection of line operators via the map~\eqref{eq:ASL2transf}.

\subsection{The standard axion-photon coupling}

Consider the theory with a dynamical axion $\theta(x)$ coupling to $F \wedge F$ for an otherwise free photon, as in~\eqref{eq:action} (with $n = 1$, for convenience). This promotes the {\em real} part of the background field $\tau$ in~\eqref{eq:tau} to a dynamical field that we sum over in the path integral, but not the imaginary part. Clearly, treating different components of $\tau$ differently in this way explicitly breaks the SL(2,$\ZZ$) duality symmetry. This is reflected in the equations of motion, in the factor of $e^2$ in the coupling of the axion to photons when the fields are canonically normalized, and in the spectrum of line and surface operators in the theory.

One way to think of the theory with an axion is that we have now {\em gauged} the $\ZZ$ subgroup of SL(2,$\ZZ$) generated by $T$, because the $T$ operation corresponds to the gauge redundancy $\theta \mapsto \theta + 2\cpi$. The $T$ operation acts trivially on the field strength $F$ (consistent with our fourth assumption in~\S\ref{sec:standard}), but it acts nontrivially on the magnetic field strength $\FM$, which shifts to $\FM + F$ (the Witten monodromy). In the theory with an axion, this means that $\FM$ is not a gauge invariant operator, even though $F$ is! This breaking of electric-magnetic duality has an important implication for the physics of magnetic charges. In particular, the 't~Hooft operator~\eqref{eq:tHooft} is no longer a genuine (gauge invariant) operator, as it transforms under $\theta \mapsto \theta + 2\cpi$. This is precisely the same issue that we discussed for physical monopoles in \S\ref{subsec:witten}, which can be resolved by introducing a localized mode $\sigma \cong \sigma + 2\cpi$ on the curve $C$ transforming under $A$ gauge transformations as~\eqref{eq:dyonicmodeshift}. As explained in~\cite{Heidenreich:2021xpr}, we can then define an 't~Hooft operator to include a path integral over this mode:
\begin{equation} \label{eq:tHooftMod}
    \widehat{T}_p(C) = \int {\cal D}\sigma \exp\left[\iu p \int_C \left(\AM - \frac{\theta}{2\cpi} (\dif \sigma + A)\right)\right].
\end{equation}
Even in the absence of dynamical monopoles, then, the anomaly inflow phenomenon manifests itself in the spectrum of line operators in the theory. Rather than the entire family of $L_{p,q}(C)$ operators, we now have Wilson operators $W_q(C)$ and 't~Hooft operators $\widehat{T}_p(C)$, but the dyonic lines have been subsumed into the 't~Hooft operator thanks to its localized degree of freedom.\footnote{One might also attempt to define $L_{p,q}(C)$ line operators for all $p$ and $q$ by imposing a $\theta = 0$ boundary condition along the line. A detailed assessment of the full range of allowed boundary conditions and line operators is beyond the scope of this work.} This reflects the breaking of SL(2,$\ZZ$) by the axion coupling.

\subsection{Alternative axion-photon couplings}
\label{subsec:alternative}

We have seen that the standard axion-photon coupling breaks SL(2,$\ZZ$) in a specific way. It gauges a specific $\ZZ$ subgroup of SL(2,$\ZZ$) corresponding to powers of $T$, which leaves Wilson line operators and magnetic 1-form surface operators untouched, while modifying 't~Hooft lines and electric 1-form surface operators. This is reflected in the effective electric, but not magnetic, charges and currents that appear in the axionic modification of Maxwell's equations~\eqref{eq:axionMaxwell}. 

Given the structure of SL(2,$\ZZ$), it is clear that this choice of axion coupling, and its privileging of electric over magnetic currents, was not a unique choice. We can define a different type of axion-photon coupling for every SL(2,$\ZZ$) element $\Lambda$~\eqref{eq:sl2z}, which defines a new duality frame. To do so, we follow a simple procedure. First, transform to the new frame; then, add a properly quantized axion coupling $\theta F \wedge F$ in the new frame; then, transform back. 

Let's begin in the frame where $A$ is the gauge field that couples to the usual electric charge carried by the electron. In this frame, we will take the complexified gauge coupling in the case where the axion field $\theta(x) = 0$ to be given by
\begin{equation}
\tau_0 = \frac{\theta_0}{2\cpi} + \iu \frac{2\cpi}{e_0^2}.
\end{equation}
The subscripts $0$ signal that these are constants, independent of field $\theta(x)$. Now, we perform an SL(2,$\ZZ$) transformation as given by~\eqref{eq:JSL2transf},~\eqref{eq:ASL2transf} and~\eqref{eq:tautransf} to the $A'$ frame. In this frame, the constant complexified gauge coupling is $\tau'_0 = (a\tau_0 + b)/(c\tau_0 + d)$. Then we add, in this frame, a new term to the action:
\begin{equation} \label{eq:modifiedaxionterm}
\delta S = \frac{k}{8\cpi^2} \int \theta(x) F'(x) \wedge F'(x).
\end{equation}
We assume that $F'$ is invariant under $\theta \mapsto \theta+ 2\cpi$, and hence $k \in \mathbb{Z}$ as derived in \S\ref{subsec:derivation}. The addition~\eqref{eq:modifiedaxionterm} changes the effective complexified gauge coupling in the $A'$ frame to 
\begin{equation}
\tau'(x) = \tau'_0 + \frac{k}{2\cpi} \theta(x).
\end{equation}
Now, we transform {\em back} to the original frame with $\Lambda^{-1}$ to find the field-dependent complexified gauge coupling
\begin{equation} \label{eq:taufullanswer}
\tau(x) = \frac{d \tau'(x) - b}{-c \tau'(x) + a} = \frac{\tau_0 + d (c\tau_0 + d)\frac{k}{2\cpi} \theta(x)}{1 - c (c\tau_0 + d)\frac{k}{2\cpi} \theta(x)}.
\end{equation}
This expression fully captures how the axion field $\theta(x)$ couples to the standard photon field:
$\mathrm{Re}\, \tau(x)$ determines the coupling to $F \wedge F$, and $\mathrm{Im}\, \tau(x)$ determines the coupling to $F \wedge \star F$. The full expressions for the real and imaginary parts of~\eqref{eq:taufullanswer} are complicated, but we can take a look at their expansion to linear order in $\theta(x)$ to see how $1/e_0^2$ and $\theta_0$ are corrected:
\begin{align} \label{eq:taulinearized}
\frac{1}{e^2(x)} \equiv \frac{1}{2\cpi} \mathrm{Im}\,\tau(x) &= \frac{1}{e_0^2} \left[1 + 2 c \left(d + c \frac{\theta_0}{2\cpi}\right) \frac{k \theta(x)}{2\cpi} + \cdots \right],  \nonumber \\
\vartheta(x) \equiv 2\cpi \mathrm{Re}\,\tau(x) &= \theta_0 + \left[\left(d + c \frac{\theta_0}{2\cpi}\right)^2 - c^2 \left(\frac{2\cpi}{e_0^2}\right)^2\right] k \theta(x) + \cdots.
\end{align}
where $\cdots$ refers to terms of order $\theta(x)^2$ or higher, and we have used the notation $\vartheta(x)$ for effective coefficient of $\frac{1}{8\cpi^2} F \wedge F$, to distinguish it from the axion field $\theta(x)$. 

\begin{figure}[!t]
\centering
\includegraphics [width = 0.9\textwidth]{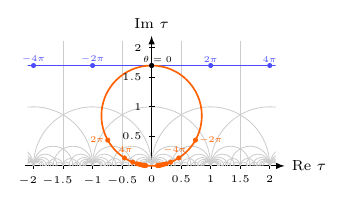}
\caption{The complexified coupling $\tau$ as a function of the axion $\theta$, as in~\eqref{eq:taufullanswer} with $\tau_0 = 1.7\iu$ and $k = 1$. The blue line is the standard $\theta F \wedge F$ axion coupling, corresponding to the choice $c = 0$, $d = 1$. In this case, the axion only affects the usual theta angle captured by $\mathrm{Re}\,\tau$. The orange curve is the case where the axion couples in an $S$-dual frame, with $d = 0$ and $c = -1$. In this case, $\tau$ asymptotically approaches zero for large values of the axion. In both cases, the dots on the curve correspond to shifts of the axion by multiples of $2\cpi$, which map $\tau_0$ to values related by an SL(2,$\ZZ$) matrix.   The faint gray lines in the background trace the boundaries of different SL(2,$\ZZ$) fundamental domains.
} \label{fig:sl2ztau}
\end{figure}
Let us highlight some key features of these results:
\begin{itemize}
\item By continuity, if $e_0^2$ is small then the effective coupling $e^2(x)$ remains small for small axion field values. However, larger axion field values $\theta(x) \sim O(1)$ can drive the coupling to be strong. 
\item When $c = 0$, the axion coupling has the expected quantized form: in this case we necessarily have $d = \pm 1$, so the coefficient is $k \in \mathbb{Z}$. 
\item When $c \neq 0$, the axion coupling in $\vartheta$ is strongly enhanced. In particular, canonical normalization multiplies the coupling by $e_0^2$, so the canonical coupling for $c \neq 0$ is proportional to $1/e_0^2$ instead of $e_0^2$ itself. This is as one would expect, from the exchange of electric and magnetic couplings under Dirac quantization.
\item When $c \neq 0$, we also observe a coupling of the axion to the standard kinetic term of the photon. Generically (unless $\theta_0 = - \frac{2\cpi d}{c} $), the coupling is linear. 
\item When $c \neq 0$, the curve $\tau(\theta)$ in the upper half-plane traced out by varying $\theta$ is a circle tangent to the real axis at $\tau = -d/c$ (a point approached asymptotically as $\theta \to \pm \infty$) and passing through the point $\tau_0$. 
\end{itemize}
We illustrate the curve $\tau(\theta)$ in two examples in Fig.~\ref{fig:sl2ztau}, starting with a purely imaginary $\tau_0$. The first is a standard axion coupling, where $\theta \mapsto \theta + 2\cpi n$ shifts $\tau \mapsto \tau + n$. The second is a coupling in an $S$-dual frame, where shifting $\theta \mapsto \theta + 2\cpi n$ acts on $\tau$ with the SL(2,$\ZZ$) transformation $S^{-1} T^n S$. In this case, $\tau(\theta)$ is a circle passing through the origin.

In general, one might expect that when a scalar field couples to $F \wedge \star F$, photon loops generate a scalar mass. This is because $F \wedge \star F$, unlike $F \wedge F$, is not a total derivative. However, the couplings~\eqref{eq:taufullanswer} are consistent with a massless axion precisely because there exists a frame in which the coupling {\em is} to a total derivative,~\eqref{eq:modifiedaxionterm}. A generic UV regulator may not respect SL(2,$\ZZ$) duality and may obscure this fact when computing loop corrections in other frames.

The results~\eqref{eq:taufullanswer} and~\eqref{eq:taulinearized} evade the argument for the axion coupling quantization given in \S\ref{subsec:derivation} for the reason anticipated in \S\ref{subsec:revisiting}: the coupling~\eqref{eq:modifiedaxionterm} is defined in a frame where $F'$ is invariant under $\theta \mapsto \theta + 2\cpi$, which means that the original field strength $F$ is {\em not} invariant under this operation. This violates the fourth assumption in the argument. Indeed, the composition of $\Lambda$ followed by $T$ followed by $\Lambda^{-1}$ corresponds to the SL(2,$\ZZ$) element
\begin{equation}
T' = \begin{pmatrix} 1 + c d & d^2 \\ -c^2 & 1 - cd \end{pmatrix},
\end{equation}
under which 
\begin{equation}   \label{eq:Ashift}
A \mapsto (1 - c d) A - c^2 \AM. 
\end{equation}
Thus, an immediate consequence of having an axion coupling that is not quantized in the standard way is that electrically charged particles acquire a magnetic charge when $\theta \mapsto \theta + 2\cpi$, in a dual form of the Witten monodromy~\eqref{eq:wittenmonodromy}. We will return to this point in \S\ref{sec:phenoassess}, but first let us discuss how our results relate to non-standard axion electrodynamics in the literature.

\subsection{Comparison to prior literature}

The expression for $\tau(x)$ in~\eqref{eq:taufullanswer} is clunky, so let's see how we can rearrange our results to more closely resemble those that have previously appeared in the literature. In the $A'$ frame, the equations of motion are simple:
\begin{align} \label{eq:primeeom}
\dif F' &= 0, \nonumber \\
\frac{1}{e'^2} \dif \star F' &= \frac{k}{4\cpi^2} \dif \theta \wedge F', \nonumber \\
f^2 \dif{\star \dif \theta} &= \frac{k}{8\cpi^2} F' \wedge F'.
\end{align}
The primed gauge coupling is independent of $\theta(x)$ and given by
\begin{equation}
e'^2 = e_0^2 \left[ c^2 \left(\frac{2\cpi}{e_0^2}\right)^2 + \left(d + c \frac{\theta_0}{2\cpi}\right)^2\right].
\end{equation}
Because these quantities will recur throughout the discussion below, it is useful to define
\begin{equation} \label{eq:deltagamma}
\gamma \equiv c \left(\frac{2\cpi}{e_0^2}\right), \quad \delta \equiv d + c \frac{\theta_0}{2\cpi},
\end{equation}
so that $e'^2 = e_0^2 \left(\gamma^2 + \delta^2\right)$.

In terms of the usual field strength $F$ and its magnetic dual $\FM$, the primed field strength is $F' = c \FM + d F$. In the special case where $\theta(x) = 0$, we further have $\FM = -\frac{2\cpi}{e_0^2} {\star F} + \frac{\theta_0}{2\cpi} F$. Let us define, in the general case, a quantity $\cF$ that is related to $F'$ in the same way that $F$ is when $\theta(x) = 0$. That is, $\cF$ is defined by
\begin{equation} \label{eq:scriptFdef}
F' = \left(d + c \frac{\theta_0}{2\cpi}\right) \cF- c \frac{2\cpi}{e_0^2} {\star \cF} = \delta \cF - \gamma {\star \cF}.
\end{equation}
Then we necessarily have $\cF \to F$ when $\theta(x) \to 0$. 

Now, we simply substitute the expression~\eqref{eq:scriptFdef} into the equations for $\dif F'$ and $\dif \star F'$ in~\eqref{eq:primeeom} and then solve for $\dif \cF$ and $\dif{\star \cF}$. We obtain:
\begin{align} \label{eq:calFequations}
\dif \star \cF + \frac{k e_0^2}{4\cpi^2} \left(-\gamma \delta  \dif \theta \wedge \star \cF + \delta^2 \dif \theta \wedge \cF\right) &= 0, \nonumber \\
\dif \cF + \frac{k e_0^2}{4\cpi^2} \left(-\gamma^2 \dif \theta \wedge \star \cF + \gamma \delta \dif \theta \wedge \cF\right) &= 0.
\end{align}
Notice that $\cF$ can't be interpreted as a field strength, because $\dif \cF \neq 0$ (even away from singular points like the core of a monopole). However, it does agree with $F$ in the limit $\theta \to 0$.

The equations~\eqref{eq:calFequations} closely resemble the equations of motion that were obtained in~\cite{Sokolov:2022fvs}. In particular, we can identify the interaction terms in our equations with the three axion-photon couplings there via
\begin{equation} \label{eq:threecouplings}
g_{AB} = \frac{k e_0^2}{4\cpi^2} \gamma \delta, \quad g_{AA} = \frac{k e_0^2}{4\cpi^2} \delta^2, \quad g_{BB} = \frac{k e_0^2}{4\cpi^2} \gamma^2,
\end{equation}
(up to normalization and sign conventions). The field strength in the original duality frame is related to $\cF$ via
\begin{equation}
F = \cF + \frac{k \theta(x)}{2\cpi} c \left(\delta \cF - \gamma {\star \cF}\right).
\end{equation}
The axion equation of motion takes the form
\begin{equation}
f^2 \dif{\star \dif \theta} =  \frac{1}{2e_0^2} \left[(g_{AA} - g_{BB}) \cF \wedge \cF - 2 g_{AB} \cF \wedge {\star \cF}\right]
\end{equation}
(where we have used that, in Minkowski signature, $\star \cF \wedge \star \cF = - \cF \wedge \cF$; compare~\eqref{eq:taulinearized}).

What we have found is that the axion couplings defined in a different SL(2,$\ZZ$) frame are essentially the new couplings of~\cite{Sokolov:2022fvs}, with the following caveats:
\begin{itemize}
\item The three couplings are not independent: in our normalization, $g_{AB}^2 = g_{AA} g_{BB}$. (However, this constraint can be relaxed for a more general coupling; see \S\ref{subsec:evenmoregeneral} and appendix~\ref{app:backwards}.)
\item The couplings obey a nontrivial quantization condition, in the sense that three integers ($c$, $d$, and $k$) fully determine their dependence on the fundamental parameters $e_0$ and $\theta_0$.
\item The field strength $\cF$ for which the equations take the simple form~\eqref{eq:calFequations} is not a field strength in the usual sense, as is clear from the fact that it is not a closed form.
\item When $c \neq 0$, the axion-photon coupling $g_{BB}$, after canonically normalizing, can be $\propto 1/e_0^2$ rather than $\propto e_0^2$. (We already noted this in the linearized analysis around~\eqref{eq:taulinearized}.) However, precisely when this large coupling appears, the electron acquires magnetic charge when $\theta \mapsto \theta + 2\cpi$.
\end{itemize}
This last point, the ``dual Witten monodromy,'' is crucial for understanding whether nonstandard axion-photon couplings are phenomenologically viable.

In Appendix~\ref{app:backwards}, we give a somewhat different and more complete perspective, starting with a set of equations of the form~\eqref{eq:calFequations} (but with completely undetermined coefficients) and systematically working out how they can map onto equations involving a closed field strength and its magnetic dual. This leads us to a very general family of functions $\tau(x)$ encoding how an axion can couple to gauge fields. Some of these couplings are simply periodic functions, where the coupling explicitly depends on $\sin(n \theta(x))$ and $\cos(n\theta(x))$. In complete theories, we expect that such couplings are suppressed by the axion mass squared, because effects that can generate such couplings can also, in general, generate an axion potential with the same spurions for violation of the continuous axion shift symmetry. We also find a set of couplings that precisely correspond to the SL(2,$\ZZ$) family of Chern-Simons couplings that we have just discussed, as well as more general couplings of the type discussed in~\S\ref{subsec:evenmoregeneral}.

\section{Phenomenological Assessment}
\label{sec:phenoassess}

In this section, we present our main arguments regarding the phenomenological viability of non-standard axion electrodynamics. We find that the dual Witten monodromy, which is implied by the presence of non-standard axion-photon couplings, is incompatible with the Standard Model, and so non-standard axion-photon couplings are phenomenologically excluded.

\subsection{The dual Witten monodromy}
\label{subsec:dualwitten}

When we couple the axion to the photon in an SL(2,$\ZZ$) dual frame, the ordinary photon field $A$ is no longer invariant under $\theta \mapsto \theta + 2\cpi$: it shifts as in~\eqref{eq:Ashift}, and in particular, acquires a term proportional to $\AM$. As a result, every electrically charged particle in the theory must acquire magnetic charge and become a dyon when $\theta \mapsto \theta + 2\cpi$. This is just the dual of the usual Witten monodromy, as explained in \S\ref{subsec:witten}. Let's begin by commenting on some general features of this dual Witten effect, before discussing it in the context of the Standard Model in particular.

\begin{figure}[!h]
\centering
\includegraphics [width = 0.5\textwidth]{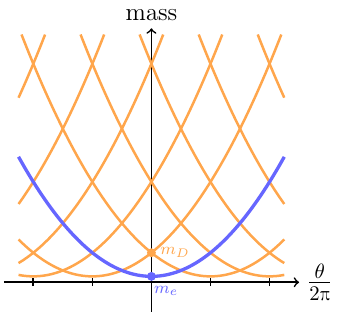}
\caption{The mass spectrum of the electron (or any other charged particle) and associated dyonic excitations, in theories with a dual Witten effect. At $\theta = 0$, the lightest state is the electron with mass $m_e$, and the first dyon appears at mass $m_D$. The blue curve tracks the mass of the electron as $\theta$ varies and it becomes a heavy dyon. The orange curves are dyonic states at $\theta = 0$; a different dyonic state plays the role of the electron at different nonzero integer values of $\theta/(2\cpi)$.
} \label{fig:monodromy}
\end{figure}

Consider an electrically charged particle, say the electron, at $\theta = 0$. If we continuously vary $\theta$ from $0$ to $2\cpi$, this particle will become a dyon. In order to have a reasonable QFT, it must be the case that its mass changes during this process. Otherwise, we would find an infinite degeneracy of dyons by tracking this state to $\theta = 2\cpi n$ for all integer $n$. Thus, the spectrum of dyonic excitations of the electron (or any other charged particle) should exhibit a monodromy, as depicted in Fig.~\ref{fig:monodromy}. The electron mass should increase as $\theta$ increases, while the mass of some other dyon state will decrease, and that state will become the new electron at $\theta = 2\cpi$. This is the same sort of behavior that we see in the context of the usual Witten effect, where the magnetic monopole mass increases as $\theta$ varies (see, e.g.,~\cite{Fischler:1983sc}).

Because the mass of the dyons depends on $\theta$, we can integrate them out to obtain an effective potential $V(\theta)$ for the axion~\cite{Fan:2021ntg}. In the usual duality frame, where the Witten effect applies to magnetic monopoles, we integrate out very heavy monopole states with small dyonic splittings. Because the monopole is a heavy semiclassical object, we should not treat it with a weakly-coupled monopole field. Instead, we include magnetic monopoles in the path integral by summing over the different paths that heavy monopole worldlines can take (see, e.g.,~\cite{Affleck:1981ag}). The sum over dyons can be recast as a sum over a winding of the dyon collective coordinate around the monopole loop, in which case the calculation admits a saddle point approximation where such monopole loops with dyonic winding can be thought of as a type of instanton~\cite{Fan:2021ntg, Heidenreich:2020pkc}.

\begin{figure}[!h]
\centering
\includegraphics [width = 0.4\textwidth]{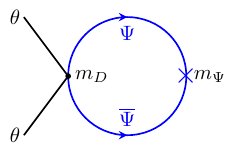}
\caption{If the electrically charged fermion $\Psi$ becomes a dyon by coupling to the axion $\theta$, then a loop of fermions can generate an axion potential. The corresponding contribution to the axion mass term can be estimated from this one-loop diagram.
} \label{fig:axionmass}
\end{figure}

When the axion couples in a non-standard duality frame, the character of our calculation changes. Now the Witten effect implies that a light, weakly-coupled particle like the electron becomes a dyon as $\theta$ varies. We treat such particles as fields in the path integral, rather than heavy semiclassical worldlines. In particular, the saddle point from the monopole calculation would now lie at small proper time and would have small action. Thus, there is no exponential suppression in the axion mass arising from electrically charged fermion loops, and the calculation lies in a regime in which we do not trust semiclassical methods. Instead, standard perturbative methods should give a reasonable estimate.\footnote{The semiclassical calculation applies when the classical radius of the charged object is much larger than its Compton radius, and standard perturbative methods apply in the opposite limit. Electrically and magnetically charged particles can never both belong to the perturbative regime.} We assume that the mass of a charged Dirac fermion $\Psi$ is approximately
\begin{equation} \label{eq:dyonmassterm}
\left[m_\Psi + m_D \left(n - \frac{\theta}{2\cpi}\right)^2 + \cdots\right] \overline{\Psi} \Psi,
\end{equation}
to quadratic order in the axion, where $n$ labels which dyon state we are considering and the $\cdots$ represent terms of higher order in $\theta$. Given such a term, and focusing on the lightest state $n = 0$ near $\theta = 0$, we estimate an axion mass from the one-loop Feynman diagram in Fig.~\ref{fig:axionmass}:
\begin{equation}
m_\theta^2 \sim \frac{1}{16\cpi^2} m_\Psi m_D \frac{\Lambda^2}{f^2} \log \frac{\Lambda}{m_\Psi}.
\end{equation}
It seems reasonable to expect that $\Lambda$ could be of order the dyon mass $m_D$, since that is a scale where new physics enters.  From this expression, we see that this contribution to the axion mass is potentially much larger than standard instanton contributions. 

So far, we have kept the discussion rather general: our only assumption is that the ``standard'' duality frame is the one in which $A$ couples to light, weakly-interacting charged particles. In this case, the standard $\theta F \wedge F$ coupling ensures that only monopole loops contribute to the axion potential, allowing for an exponentially small (semiclassical) axion mass. In any other frame, where the coupling takes the form $\theta F' \wedge F'$, we expect the axion to obtain a large mass.

Now, let's turn to a more realistic axion phenomenology, with the axion coupled to the Standard Model photon. Every charged particle contributes in loops like Fig.~\ref{fig:axionmass}. In particular, the top quark does. Taking $m_\Psi = m_t$ and taking $m_D \sim \Lambda \sim 1\,\mathrm{TeV}$, we immediately see that a top quark loop contributes an axion mass
\begin{equation}
m_\theta \sim 30\,\mathrm{eV}\, \frac{10^{12}\,\mathrm{GeV}}{f}.
\end{equation}
This completely overwhelms the standard axion mass. There is no possibility of suppressing $m_D$ or $\Lambda$, since we have already probed physics up to the TeV scale.  There is no a priori reason for these contributions to the axion potential to have the same phase as the QCD contribution, so this would spoil the solution to the Strong CP problem.

In fact, the situation is even worse than this. The Standard Model fermions are chiral and weakly interacting, and we assume that their interactions with the axion are also weak (a reasonable assumption, if we are discussing anything resembling standard axion phenomenology). Thus, not only the usual fermion mass $m_\Psi$ but also the dyon mass term $m_D$ in~\eqref{eq:dyonmassterm} should be proportional to the Higgs vev. It does not make sense to imagine that the electron acquires a mass in a $\theta$ background in the limit that the Higgs field has no expectation value. This immediately tells us that $m_D$ should not be far above the TeV scale, because there is a unitarity bound on the strength of the interaction of the Higgs with the fermion~\cite{Itzykson:1980rh}. In fact, the {\em entire tower} of dyons should have this property! In the limit that the Higgs vev is turned off, the electron should not acquire a mass as $\theta$ varies, and we can vary it over many cycles to find that every dyon becomes massless. This is a complete disaster for field theory, with an infinite tower of massless states.

Although the nonzero Higgs vev means that our universe does not strictly reside in this limit, the theory nonetheless predicts a tower of particles, all with mass tied to the electroweak scale and interacting with the Higgs. These particles run in loops in processes involving the Higgs boson, and it is difficult to see how the theory could remain weakly coupled in any sense. Concretely, all of these dyons appear in triangle diagrams contributing to the Higgs couplings to photons and (for dyonic excitations of quarks) to gluons, which are empirically known to be approximately the values predicted by the Standard Model. These measurements, independent of any speculation about how a viable model of strongly-coupled QFT could accommodate all of the dyon states, are sufficient to phenomenologically exclude such a model.

We expect that a stronger statement is true, that such a theory (with an infinite tower of dyons obtaining mass via the Higgs mechanism) is simply inconsistent. In recent years there has been intensive study of theories in which infinite towers of particles become massless at a point in scalar field space (e.g.,~\cite{Ooguri:2006in, Heidenreich:2018kpg, Grimm:2018ohb, Lee:2019wij, Lanza:2021udy, Stout:2021ubb, Stout:2022phm, Etheredge:2022opl}). When the particles can be treated as approximately elementary, the loop effects of these particles modify the scalar kinetic term and make the scalar $\to 0$ limit an infinite distance limit~\cite{Heidenreich:2018kpg, Grimm:2018ohb}. In the case we are discussing, this scalar would be the Higgs. Such infinite distance limits are believed to happen only in weak-coupling limits of quantum gravity, in which the scalar field controlling the tower's mass in all known limits parametrizes either the volume of decompactifying extra dimensions or the tension of an emergent light string~\cite{Lee:2019wij}. None of these examples resemble the Higgs boson; for example, none of the scalar fields carry nonabelian gauge charge. There is another alternative: when the states in the tower are strongly interacting (as the dyons are expected to be), the origin of field space may not lie at infinite distance, but instead may be a strongly interacting CFT.\footnote{See, e.g.,~\cite{Witten:1996qb} for examples of this phenomenon in 5d, which should lead to similar 4d examples upon dimensional reduction.} In examples, the gauge theory under which the particles in the tower are charged is emergent, and does not exist at the origin of field space. It is unclear if a Standard Model-like theory could ever arise by perturbing such a theory onto a Higgs branch. Perhaps the most general argument we can give is that integrating out a tower of particles of increasingly large charge under a gauge symmetry is generically expected to drive that gauge theory to weak coupling (e.g.,~\cite{Heidenreich:2017sim}). For the dyon tower, it is the magnetic charge that grows as we ascend the tower, and so we would expect the {\em magnetic} coupling to be driven small. Accordingly, the electric coupling would become {\em large}---the opposite of what we see in the Standard Model. However, we emphasize that even if our doubts are ill-founded and a theory with chiral fermions accompanied by dyonic towers actually exists, it is not phenomenologically viable for the reasons discussed above; none of the conclusions of this paper rest on this paragraph.

Our arguments have focused on chiral fermions obtaining a mass from a Higgs, as in the Standard Model, but one might ask (and a referee has): could a problem with non-standard axion electrodynamics exist already in the more general case of a charged Dirac fermion whose mass $m$ goes to zero? For example, could one conjecture that an axion coupling $\theta F \wedge F$ is compatible only with massless fermions electrically charged under $A$, not dyons charged under a combination of $A$ and $\AM$? (Then, correspondingly, a non-standard axion coupling would be incompatible with purely electrically charged massless fermions, and we would not need to appeal to Higgs physics.) In the precise form we have asked the question, this is false. A counterexample is provided by Seiberg-Witten theory~\cite{Seiberg:1994rs}, which can be weakly coupled to a holomorphic modulus $\tau$ (including, in particular, an axion coupling to $F \wedge F$ for the low energy $\Uone$ gauge theory), and which has a point in its moduli space at which a magnetic monopole becomes massless. In fact, at this point in moduli space, the magnetic coupling becomes arbitrarily weak in the IR in the $m \to 0$ limit, due to the usual logarithmic running in QED. Thus, any more general no-go theorem that does not rely on chiral fermions would need to rely on some additional assumptions.

\subsection{More general couplings}
\label{subsec:evenmoregeneral}

In our discussion so far, we have focused on axion couplings that take the standard form $\theta F \wedge F$ in some duality frame. This corresponds to the requirement that the gauge transformation $\theta \mapsto \theta + 2\cpi$ acts on the complex coupling parameter $\tau$ via an SL(2,$\ZZ$) element of the form $\Lambda^{-1} T^n \Lambda$. (These are known as the parabolic elements of SL(2,$\ZZ$), those with the absolute value of the trace equal to 2.) One could also consider an even more general axion coupling: given any SL(2,$\ZZ$) element $\Lambda = \begin{pmatrix} a & b \\ c & d \end{pmatrix}$, we could consider {\em any} function $\tau(\theta)$ with a $\Lambda$-twisted periodicity property, i.e.,
\begin{equation}
\tau(\theta + 2\cpi) = \frac{a\tau(\theta)+b}{c\tau(\theta)+d}.
\end{equation}    
Then $\theta \mapsto \theta  + 2\cpi$ returns the theory to itself up to a duality transformation. Here we will offer some brief remarks about these more general possibilities.

The simplest examples of such more general functions are those that correspond to the duality frames we have already discussed, but with additional periodic contributions to $\tau(\theta)$. For instance, we might have a coupling of the form $\sin(\theta) F \wedge F$. Because this is manifestly gauge invariant, it can come with an arbitrary coupling constant. Such couplings arise in ordinary quantum field theories, and are generally expected to be suppressed by the mass of the axion squared, because physics that generates such a coupling could also generate a periodic potential $V(\theta)$. An interesting, well-known, example is the contribution to the axion-photon coupling arising from the axion mixing with the pion. This contribution is larger than one might naively expect, because the axion and pion masses both arise from QCD dynamics~\cite{Kaplan:1985dv, Srednicki:1985xd, Georgi:1986df, Svrcek:2006yi, Agrawal:2017cmd, Agrawal:2023sbp}. It gives an $O(1)$ contribution $\delta n$ modifying the prefactor in the coupling~\eqref{eq:gagg}. In the KSVZ model~\cite{Kim:1979if,Shifman:1979if}, where $n = 0$ in~\eqref{eq:action}, the pion mixing generates the only axion-photon coupling and we are in the $\Lambda = 1$ case. In other models, like the DFSZ model~\cite{Dine:1981rt,Zhitnitsky:1980tq}, where $n \neq 0$, we have $\Lambda = T^n$ but now the function $\tau(\theta)$ is the sum of a term linear in $\theta$ and a term periodic in $\theta$.

Now, consider the most general case. The only SL(2,$\ZZ$) elements under which electrically charged particles do not acquire magnetic charge are those of the form $T^n$, corresponding to standard axion couplings, or $-T^n$, in which case the $\theta \mapsto \theta + 2\cpi$ operation is accompanied by charge conjugation. Any other choice, then, will imply that the electrically charged particles of the Standard Model have a family of dyonic excitations, with associated phenomenological difficulties. If $\Lambda$ is an element of infinite order (either the parabolic type already discussed, or a hyperbolic element with absolute value of the trace $> 2$), then we have an infinite tower of dyon modes, and the theory is pathological for the reasons discussed in \S\ref{subsec:dualwitten}. One more interesting possibility remains: $\Lambda$ could be a nontrivial SL(2,$\ZZ$) element of finite order (an elliptic element, with absolute value of the trace $< 2$). Apart from the $\ZZ_2$ subgroup generated by charge conjugation, which is not of interest to us, SL(2,$\ZZ$) has finite subgroups isomorphic to $\ZZ_3$, $\ZZ_4$, and $\ZZ_{6}$. For example, $S$ itself is an element of order 4, while $ST$ has order 6. In such cases, one would have only a finite number of dyonic excitations of charged particles.

Could the Standard Model be coupled to an axion with such a finite monodromy orbit? We believe that this is again problematic, though less pathological than the case with an infinite tower of dyons. Some of the arguments of \S\ref{subsec:dualwitten} continue to apply: dyon loops would again generate large corrections to the axion potential. In fact, because there is no frame in which these generalized couplings take the form $\theta F \wedge F$ in which the axion couples to a total derivative, we would expect that {\em photon} loops modify the axion potential as well. We would again expect that the dyonic partners of Standard Model fermions can only obtain a mass from electroweak symmetry breaking, so they would have mass near the TeV scale, and would also alter the Higgs couplings to photons and gluons away from their Standard Model predictions. (If such a theory exists and could be reconciled with precision Higgs physics, it would provide a novel motivation for searches for monopoles and dyons at the TeV scale, like~\cite{MoEDAL:2016lxh, MoEDAL:2020pyb}.) From a more theoretical viewpoint, one should take care that the SL(2,$\ZZ$) elements that act on the theory are not anomalous~\cite{Witten:1995gf, Hsieh:2019iba}. Another problem, when $\Lambda$ is of even order, is that a power of $\Lambda$ corresponds to charge conjugation, which is not a symmetry of the Standard Model. One would need more elaborate model-building to make sense of this. Finally, it is not at all clear what form a UV completion of such a coupling could take. Because the coupling $\tau(\theta)$ is a periodic function in this case, we would tend to expect the coefficients of such couplings to be highly suppressed, for similar reasons to the standard periodic couplings mentioned above. A case in which $\theta \mapsto \theta+ 2\cpi$ generates a finite monodromy could be thought of as a coupling generated by a novel sort of ``fractional instanton,'' and would be expected to have an exponentially small coefficient. All of these considerations make it highly unlikely that a consistent theory of an axion coupled to the Standard Model with a finite monodromy orbit could exist. 

\section{Conclusions}
\label{sec:conclusions}

It is a familiar fact about a wide variety of axion theories that the axion coupling to photons is quantized in units of $e^2/(8\cpi^2 f)$, when the fields are canonically normalized. Recently this conventional wisdom has been called into question, especially by~\cite{Sokolov:2022fvs}. In agreement with that work, we find that a wider variety of axion couplings to $\Uone$ gauge theory are possible. These correspond to the possibility that $\theta \mapsto \theta + 2\cpi$ is accompanied by a nontrivial SL(2, $\ZZ$) electromagnetic duality transformation. However, we conclude that these non-standard theories of axion electrodynamics are incompatible with the real world, due to the existence of electrically charged chiral fermions in the Standard Model, which would acquire dyonic excitations if such non-standard axion couplings exist. This is inconsistent with the Standard Model as a weakly coupled effective field theory in which electroweak symmetry is broken only by the Higgs boson, as indicated by experimental results.

The Witten effect, and in particular the monodromy of the spectrum of charged objects that arises under $\theta \mapsto \theta + 2\cpi$, played a key role in our discussion. We reviewed a simple argument for the inevitability of the Witten monodromy in \S\ref{subsec:witten}. Standard axion electrodynamics can be thought of as gauging the $\ZZ$ subgroup of SL(2,$\ZZ$) generated by $T$. One class of non-standard couplings can be thought of as instead gauging the $\ZZ$ subgroup whose elements are powers of the element $\Lambda^{-1} T \Lambda \in \mathrm{SL}(2,\ZZ)$. The spectrum still undergoes a monodromy under this $\ZZ$ subgroup, but for nontrivial $\Lambda$, electrically charged particles are part of a tower of dyons carrying magnetic charge. Another class of non-standard couplings has only a finite monodromy orbit, but still implies that electrically charged particles become dyons as the axion field value varies. All of these possibilities are excluded by Higgs physics.

As mentioned in \S\ref{subsec:evenmoregeneral}, the quantization of the axion-photon coupling applies only to the Chern-Simons coupling, not to additional couplings like $\sin(\theta)F \wedge F$ that are manifestly gauge invariant. An important such contribution arises from the QCD axion's mixing with the pion. There are some subtleties in the Chern-Simons couplings themselves. First, the quantization rule depends on the basic quantum of $\Uone$ charge; if we discovered a particle of hypercharge $1/12$, for instance, our conclusion about the allowed base unit of the axion-photon coupling would change. In the Standard Model, an additional subtlety arises from the global structure of the gauge group, which is ambiguous since there are elements of the center of $\SU(2)_\textsc{L}$ and $\SU(3)_\textsc{C}$ that act on all known fields in the same way as elements of $\Uone_\textsc{Y}$~\cite{Tong:2017oea}. This allows the existence of field configurations with correlated fractional topological charges~\cite{tHooft:1979rtg,vanBaal:1982ag,Anber:2021upc}, which lead to quantization rules that correlate the axion couplings to gluons and photons~\cite{Choi:2023pdp,Reece:2023iqn,Cordova:2023her}. In general non-abelian gauge theories, one can also modify the path integral to include only field configurations with topological charge a multiple of some base unit $p \neq 1$~\cite{Pantev:2005rh, Pantev:2005wj, Pantev:2005zs, Seiberg:2010qd, Tanizaki:2019rbk}. 

Finally, even more exotic generalized axion couplings are known to arise in various examples, such as Kaluza-Klein reduction of 5d gauge theory. In this case, we find a coupling of the form $\theta^3 H \wedge H$ of an axion $\theta$ to the KK field strength $H$, which is not invariant under $\theta \mapsto \theta + 2\cpi$. However, it appears as part of a monodromy with a different gauge field that has field strength $F$, in a structure of the schematic form $\theta F \wedge F + \theta^2 H \wedge F+ \theta^3 H \wedge H$ (with appropriate coefficients). Under $\theta \mapsto \theta + 2\cpi$, we have $F \mapsto F - H$, which ensures consistency of the whole structure. Such generalized theta terms have recently been examined in~\cite{Heidenreich:2021yda, Grimm:2022xmj}. These examples are qualitatively similar to the SL(2,$\ZZ$) alternatives we have discussed in this paper, in the sense that they rely on gauge field strengths that transform nontrivially under $2\cpi$ shifts of the axion. From the phenomenological standpoint, they lead to weaker axion interactions than the standard couplings, so they do not seem to pose an interesting loophole. 

The physics of axion-photon couplings is very rich, with a number of subtleties and interesting applications of topology in quantum field theory. Nonetheless, the equations of axion electrodynamics~\eqref{eq:axionMaxwell}, presented by Sikivie already forty years ago~\cite{Sikivie:1983ip}, are the correct equations that should guide experimental searches for an axion or axion-like particle coupling to photons.

\section*{Acknowledgments}
MR thanks Prateek Agrawal and John Terning for raising thought-provoking questions (in conversations a few years ago) about how electric-magnetic duality relates to axion physics, and Kevin Zhou for providing references to some of the literature on non-standard formulations of axion electrodynamics. MR also thanks Anton Sokolov for an email exchange, and Eduardo Garc{\'i}a-Valdecasas for comments on a draft. We thank an anonymous referee for questions and comments that have improved the paper. BH is supported by NSF grant PHY-2112800. JM is supported by the U.S. Department of Energy, Office of
Science, Office of High Energy Physics, under Award Number DE-SC0011632. MR is supported in part by the DOE Grant DE-SC0013607 and the NASA Grant 80NSSC20K0506. 

\appendix 

\section{Working backwards: equations of motion to a standard action}
\label{app:backwards}

In the main text, we started with a manifestly well-defined axion coupling in some choice of SL(2,$\ZZ$) duality frame, and then showed that an appropriate definition of a non-standard field strength $\cF$ could recast the equations of motion in the form~\eqref{eq:calFequations} that was studied in \cite{Sokolov:2022fvs}. In this appendix, we work in the other direction. Beginning with a hypothesized set of equations of motion for $\cF$ coupled to a shift-symmetric scalar $\phi$, we re-express them in terms of a standard field strength $F$ with complexified gauge coupling determined by $\phi$. The latter admits a standard quantization via a generalized Maxwell action, allowing us to determine the quantization conditions on the axion-photon couplings.

\subsection{Relating the Sokolov-Ringwald equations to standard
electrodynamics}

Motivated by \cite{Sokolov:2022fvs}, we consider classical electrodynamic equations of
the form:\footnote{These are related to the equations in \cite{Sokolov:2022fvs} via
$\bigl(\begin{smallmatrix}
  g_{11} & g_{12}\\
  g_{21} & g_{22}
\end{smallmatrix}\bigr)^{(\text{here})} = \bigl(\begin{smallmatrix}
  g_{a A B} & - g_{a A A}\\
  g_{a B B} & - g_{a A B}
\end{smallmatrix}\bigr)^{(\text{there})}$.}
\begin{equation}
\begin{aligned}
  \dif\star \cF+ \dif\phi \wedge (g_{11} \star \cF-
  g_{12} \cF) &= 0,  \\
  \dif\cF+ \dif\phi \wedge (- g_{21} \star \cF+ g_{22}
  \cF) &= 0 .  
\end{aligned} \label{eqn:SReqns}
\end{equation}
Note that we can set $g_{22} = - g_{11}$ after a field redefinition
$\cF \rightarrow \exp \bigl[ - \frac{g_{11} + g_{22}}{2} \phi \bigr]
\cF$, so we assume this to be the case henceforward.

These equations have the virtue that they are invariant under constant axion
shifts, $\phi \rightarrow \phi + \delta \phi$. However, since the
field-strength tensor $\cF$ is not closed, we cannot introduce a gauge
potential $\cA$ such that $\cF= \dif \cA$ in the
standard way, making quantization difficult. Instead of using the Zwanziger
formalism as in \cite{Sokolov:2022fvs}, we aim to rewrite
these equations in standard form via a field redefinition, i.e., we seek
functions $F (\cF, \phi)$, $e (\phi)$ and $\theta (\phi)$ such that
(\ref{eqn:SReqns}) becomes:
\begin{equation}
  \dif F = 0, \qquad \dif \FM = 0, \qquad \text{where} \qquad \FM \equiv -
  \frac{2 \cpi}{e^2 (\phi)} \star F + \frac{\theta (\phi)}{2 \cpi} F.
\end{equation}
Here $F = \dif A$ is a standard $U (1)$ gauge field with gauge coupling $e
(\phi)$ and theta angle $\theta (\phi)$, whose quantization is well known (see \S\ref{sec:standard}, \S\ref{sec:magnetic}).

To do so, let us assume that $F|_{\phi = 0} =\cF$.\footnote{More
generally, if $F|_{\phi = 0} = \alpha \cF+ \beta \star \cF$ for
constants $\alpha, \beta$ then we first rewrite (\ref{eqn:SReqns}) in terms of
$\cF' = \alpha \cF+ \beta \star \cF$:
$$ \begin{aligned}
     \dif \star \cF' + \dif \phi \wedge (g_{11}' \star
     \cF' - g_{12}' \cF') &= 0,\\
     \dif \cF' + \dif \phi \wedge (- g_{21}' \star \cF' +
     g_{22}' \cF') &= 0,
   \end{aligned} \qquad \text{where} \qquad \begin{pmatrix}
     g_{11}' & g_{12}'\\
     g_{21}' & g_{22}'
   \end{pmatrix} = \begin{pmatrix}
     \alpha & \beta\\
     - \beta & \alpha
   \end{pmatrix} \begin{pmatrix}
     g_{11} & g_{12}\\
     g_{21} & g_{22}
   \end{pmatrix} \begin{pmatrix}
     \alpha & \beta\\
     - \beta & \alpha
   \end{pmatrix}^{- 1} . $$} With foresight, we
first define
\begin{equation}
  \cFM \equiv - \frac{2 \cpi}{e^2_0} \star \cF+
  \frac{\theta_0}{2 \cpi} \cF, \qquad \text{where} \qquad e_0 = e (0),
  \qquad \phi_0 = \phi (0) . \label{eqn:cFMdefn}
\end{equation}
In terms of $\cF, \cFM$, (\ref{eqn:SReqns}) becomes:
\begin{equation}
  \begin{aligned}
    \dif \cFM + \dif \phi \wedge (k_{11} \cFM + k_{12}
    \cF) &= 0,\\
    \dif \cF+ \dif \phi \wedge (k_{21} \cFM + k_{22}
    \cF) &= 0,
  \end{aligned} \quad \text{where} \quad \begin{pmatrix}
    k_{11} & k_{12}\\
    k_{21} & k_{22}
  \end{pmatrix} = \begin{pmatrix}
    \frac{2 \cpi}{e^2_0} & \frac{\theta_0}{2 \cpi}\\
    0 & 1
  \end{pmatrix} \! \begin{pmatrix}
    g_{11} & g_{12}\\
    g_{21} & g_{22}
  \end{pmatrix} \! \begin{pmatrix}
    \frac{2 \cpi}{e^2_0} & \frac{\theta_0}{2 \cpi}\\
    0 & 1
  \end{pmatrix}^{- 1}, \label{eqn:SReqns2}
\end{equation}
and $k_{22} = - k_{11}$. Thus, by construction $F|_{\phi = 0} =\cF$ and $\FM|_{\phi = 0}
=\cFM$. More generally, for $\phi \neq 0$:
\begin{equation}
  \begin{pmatrix}
    \FM\\
    F
  \end{pmatrix} = \begin{pmatrix}
    a (\phi) & b (\phi)\\
    c (\phi) & d (\phi)
  \end{pmatrix} \begin{pmatrix}
    \cFM\\
    \cF
  \end{pmatrix}, \qquad \text{where} \qquad \left.
  \begin{pmatrix}
    a & b\\
    c & d
  \end{pmatrix} \right|_{\phi = 0} = \begin{pmatrix}
    1 & 0\\
    0 & 1
  \end{pmatrix} . \label{eqn:Fredef}
\end{equation}
To determine the functions $a (\phi), b (\phi), c (\phi), d (\phi)$, we impose
Maxwell's equations $\dif F = \dif \FM = 0$ and apply (\ref{eqn:SReqns2}).
This yields the differential equation
\begin{equation}
  \frac{\rmd}{\rmd \phi}\!  \begin{pmatrix}
    a & b\\
    c & d
  \end{pmatrix} = \begin{pmatrix}
    a & b\\
    c & d
  \end{pmatrix} \! \begin{pmatrix}
    k_{11} & k_{12}\\
    k_{21} & k_{22}
  \end{pmatrix} , \qquad \text{so that} \qquad \begin{pmatrix}
    a & b\\
    c & d
  \end{pmatrix} = \exp \biggl[ \phi \begin{pmatrix}
    k_{11} & k_{12}\\
    k_{21} & k_{22}
  \end{pmatrix} \biggr] , \label{eqn:SL2Rmatrix}
\end{equation}
upon imposing the $\phi = 0$ boundary condition (\ref{eqn:Fredef}).
Using \eqref{eqn:cFMdefn}, \eqref{eqn:Fredef}, we obtain
\begin{equation}
  \FM = - a \frac{2 \cpi}{e^2_0} \star \cF+ \biggl( a \frac{\theta_0}{2
  \cpi} + b \biggr) \cF, \qquad F = - c \frac{2 \cpi}{e^2_0} \star
  \cF+ \biggl( c \frac{\theta_0}{2 \cpi} + d \biggr) \cF.
\end{equation}
Solving the second equation for $\cF$ and substituting into the first,
one finds after some algebra that
\begin{equation}
  \FM = - \Im \biggl( \frac{a \tau_0 + b}{c \tau_0 + d} \biggr) \star F +
  \Re \biggl( \frac{a \tau_0 + b}{c \tau_0 + d} \biggr) F, \qquad
  \text{where} \qquad \tau_0 \equiv \frac{\theta_0}{2 \cpi} + \iu \frac{2
  \cpi}{e_0^2} .
\end{equation}
Thus, the axion-dependent coupling constants are given by
\begin{equation}
  \tau (\phi) \equiv \frac{\theta (\phi)}{2 \cpi} + \frac{2 \cpi \iu}{e (\phi)^2}
  = \frac{a (\phi) \tau_0 + b (\phi)}{c (\phi) \tau_0 + d (\phi)} .
\end{equation}
This is simply a $\PSL(2, \mathbb{R})$ transformation of the $\phi = 0$
coupling $\tau_0$ by $\bigl(\begin{smallmatrix}
  a & b\\
  c & d
\end{smallmatrix}\bigr) = \exp \bigl[ \phi \bigl(\begin{smallmatrix}
  k_{11} & k_{12}\\
  k_{21} & k_{22}
\end{smallmatrix}\bigr) \bigr]$.

We can now write an action leading to \eqref{eqn:SReqns}:
\begin{multline}
  S = - \frac{(2 \cpi f)^2}{2} \! \int \dif \phi \wedge \star \dif \phi +
  \frac{1}{4 \cpi} \int F \wedge [\Re \tau (\phi) F - \Im \tau
  (\phi) \star F], \\
  \text{where} \qquad F = \dif A, \qquad \tau (\phi) = \frac{a \tau_0 + b}{c \tau_0
  + d}, \qquad \begin{pmatrix}
    a & b\\
    c & d
  \end{pmatrix} = \exp \biggl[ \phi \begin{pmatrix}
    k_{11} & k_{12}\\
    k_{21} & - k_{11}
  \end{pmatrix} \biggr] . \label{eqn:StdAction} 
\end{multline}
Here $f$ is the axion decay constant, $\tau_0 = \frac{\theta_0}{2 \cpi} + \iu
\frac{2 \cpi}{e_0^2}$ is the complexified gauge coupling at $\phi = 0$, and
$k_{11}, k_{12}, k_{21}$ are additional real constants. Defining $\cF
\equiv \Re \bigl[ \frac{F + \iu \star F}{c \tau_0 + d} \bigr]$ and
following the same steps as above in reverse, one recovers \eqref{eqn:SReqns}
with
\begin{equation}
  \begin{pmatrix}
    g_{11} & g_{12}\\
    g_{21} & g_{22}
  \end{pmatrix} = \begin{pmatrix}
    \frac{2 \cpi}{e^2_0} & \frac{\theta_0}{2 \cpi}\\
    0 & 1
  \end{pmatrix}^{- 1} \! \begin{pmatrix}
    k_{11} & k_{12}\\
    k_{21} & - k_{11}
  \end{pmatrix} \! \begin{pmatrix}
    \frac{2 \cpi}{e^2_0} & \frac{\theta_0}{2 \cpi}\\
    0 & 1
  \end{pmatrix} .
\end{equation}
To complete the comparison with \cite{Sokolov:2022fvs}, we consider the axion equation
of motion:
\begin{equation}
  (2 \cpi f)^2 \dif \star \dif \phi = \frac{1}{4 \cpi} F \wedge [- \Re
  \tau' (\phi) F + \Im \tau' (\phi) \star F] .
\end{equation}
Re-expressing this in terms of $\cF$ using the relation $\cF+
\iu \star \cF= \frac{F + \iu \star F}{c \tau_0 + d}$ and applying
\begin{equation}
  (c \tau_0 + d)^2 \tau' (\phi) = 2 k_{11} \tau_0 + k_{12} - \tau_0^2 k_{21} =
  (g_{21} + g_{12} + 2 \iu g_{11}) \Im \tau_0,
\end{equation}
we find that
\begin{equation}
  (2 \cpi f e_0)^2 \dif \star \dif \phi = - \frac{g_{21} + g_{12}}{2}
  \cF \wedge \cF+ g_{11} \cF \wedge \star \cF.
  \label{eqn:SRaxion}
\end{equation}
This matches with \cite{Sokolov:2022fvs} up to signs.

Note that in the special case $g_{11} = 0$, $g_{21} = - g_{12}$, $\cF$
decouples from the axion equation of motion \eqref{eqn:SRaxion}. To understand 
why, note that in this case 
\begin{align}
  \begin{pmatrix}
    a & b\\
    c & d
  \end{pmatrix} 
  &= \begin{pmatrix}
    \cos (g_{12} \phi) - \frac{\Re \tau_0}{\Im \tau_0} \sin
    (g_{12} \phi) & \frac{| \tau_0 |^2}{\Im \tau_0} \sin (g_{12} \phi)\\
    - \frac{1}{\Im \tau_0} \sin (g_{12} \phi) & \cos (g_{12} \phi) +
    \frac{\Re \tau_0}{\Im \tau_0} \sin (g_{12} \phi)
  \end{pmatrix} ,
\end{align}
using \eqref{eqn:SReqns2}, \eqref{eqn:SL2Rmatrix}, so that
\begin{equation}
  \tau (\phi) = \frac{\bigl[ \cos (g_{12} \phi) - \frac{\Re
  \tau_0}{\Im \tau_0} \sin (g_{12} \phi) \bigr] \tau_0 + \frac{| \tau_0
  |^2}{\Im \tau_0} \sin (g_{12} \phi)}{- \frac{1}{\Im \tau_0} \sin
  (g_{12} \phi) \tau_0 + \cos (g_{12} \phi) + \frac{\Re
  \tau_0}{\Im \tau_0} \sin (g_{12} \phi)} = \frac{\cos (g_{12} \phi) - \iu
  \sin (g_{12} \phi)}{\cos (g_{12} \phi) - \iu \sin (g_{12} \phi)} \tau_0 =
  \tau_0 .
\end{equation}
As a result $\phi$ and $A$ decouple from each other in
\eqref{eqn:StdAction}.\footnote{Although $\phi$ still appears in
\eqref{eqn:SReqns}, it can be removed by the field redefinition $\cF
\rightarrow \cos (g_{12} \phi) \cF- \sin (g_{12} \phi) \star
\cF$. This explains why \eqref{eqn:SRaxion} only
depends on the combination $g_{12} + g_{21}$. In \cite{Sokolov:2022fvs}, the
combination $g_{12} - g_{21}$ appears instead, likely as the result of a typo or sign error
somewhere.}\textsuperscript{,\,}\footnote{If $\tau_0$ is invariant under a non-trivial element $\Lambda_1 \in \SL(2,
\mathbb{Z})$ then the interesting possibility remains that $\phi \rightarrow \phi + 1$ acts
non-trivially on $\bigl(\begin{smallmatrix}
  \FM\\
  F
\end{smallmatrix}\bigr)$ while leaving $\tau$ unchanged. 
However, this can only occur at
strong coupling $e_0 \geqslant \sqrt{2 \cpi}$, so it is not phenomenologically
relevant.}

\subsection{Quantization of the generalized axion couplings}

So far, we have shown that the generalized axion electrodynamics equations
derived in \cite{Sokolov:2022fvs} follow from a standard action \eqref{eqn:StdAction}
at the classical level (up to a likely sign error in the axion equation of
motion given in \cite{Sokolov:2022fvs}). The quantization of \eqref{eqn:StdAction} is
straightforward, along the lines discussed in \S\ref{sec:standard}, \S\ref{sec:magnetic}. In
particular, given the assumptions discussed in \S\ref{subsec:revisiting}, we are
interested in the case where $F$ is a (holomorphically normalized) $U (1)$
gauge field and $\phi \cong \phi + 1$ is a compact scalar. Then we are forced
to impose the consistency condition that the monodromy matrix lies within $\SL(2,\mathbb{Z})$:
\begin{equation}
  \Lambda_1 \equiv \left. \begin{pmatrix}
    a & b\\
    c & d
  \end{pmatrix} \right|_{\phi = 1} = \exp \biggl[ \begin{pmatrix}
    k_{11} & k_{12}\\
    k_{21} & - k_{11}
  \end{pmatrix} \biggr] \in \SL(2, \mathbb{Z}),
  \label{eqn:phiPeriod}
\end{equation}
so that the shift symmetry $\phi \cong \phi + 1$ is exact.

Written out explicitly, the precise form of the constraint
\eqref{eqn:phiPeriod} depends on the sign of $\vartheta^2 \equiv k_{11}^2 +
k_{12} k_{21}$. We first consider the case where $\vartheta^2 > 0$, for which we
obtain:
\begin{equation}
  \Lambda_1 = \begin{pmatrix}
    \cosh \vartheta + k_{11}  \frac{\sinh \vartheta}{\vartheta} & k_{12}  \frac{\sinh \vartheta}{\vartheta}\\
    k_{21}  \frac{\sinh \vartheta}{\vartheta} & \cosh \vartheta - k_{11}  \frac{\sinh \vartheta}{\vartheta}
  \end{pmatrix} \in \SL(2, \mathbb{Z}) .
\end{equation}
Taking the trace, we conclude that $\cosh \vartheta = \frac{n}{2}$ for $n \in
\mathbb{Z}$, $n > 2$. A general solution then takes the form $k_{i j} = \frac{\vartheta}{2 \sinh \vartheta} n_{i j}$
for integers $n_{ij}$ satisfying $n_{11}^2 + n_{12} n_{21} = n^2 - 4$ with $n_{12}$ and $n_{21}$ even.

In other words, given a monodromy matrix
  $\bigl(\begin{smallmatrix}
    a_1 & b_1\\
    c_1 & d_1
  \end{smallmatrix}\bigr) \in \SL(2, \mathbb{Z})$ with trace $a_1 + d_1 > 2$,
the couplings $k_{i j}$ are fixed to be
\begin{equation}
  \begin{pmatrix}
    k_{11} & k_{12}\\
    k_{21} & - k_{11}
  \end{pmatrix} = \frac{\cosh^{- 1} \bigl( \frac{a_1 + d_1}{2}
  \bigr)}{\sqrt{\bigl( \frac{a_1 + d_1}{2} \bigr)^2 - 1}} 
  \begin{pmatrix}
    \frac{a_1 - d_1}{2} & b_1\\
    c_1 & \frac{d_1 - a_1}{2}
  \end{pmatrix}, \label{eqn:kLambdaEqn}
\end{equation}
so the choice of a monodromy matrix $\Lambda_1\in \SL(2,\mathbb{Z})$ with trace $\Tr \Lambda_1 > 2$ fully
fixes the couplings.

Next, consider the case $\vartheta^2 = 0$, for which
\begin{equation}
  \Lambda_1 = \begin{pmatrix}
    1 + k_{11} & k_{12}\\
    k_{21} & 1 - k_{11}
  \end{pmatrix} \in \SL (2, \mathbb{Z}),
\end{equation}
so that $k_{i j} \in \mathbb{Z}$ with $k_{11}^2 + k_{12} k_{21} = 0$. As
before, this implies that the couplings $k_{i j}$ are fully fixed by the
monodromy matrix $\Lambda_1$ when $\Tr \Lambda_1 = 2$:
\begin{equation}
  \begin{pmatrix}
    k_{11} & k_{12}\\
    k_{21} & - k_{11}
  \end{pmatrix} = \begin{pmatrix}
    \frac{a_1 - d_1}{2} & b_1\\
    c_1 & \frac{d_1 - a_1}{2}
  \end{pmatrix}, \label{eqn:kLambdaEqn2}
\end{equation}
except that the special case $\Lambda_1 = \bigl(\begin{smallmatrix}
  1 & 0\\
  0 & 1
\end{smallmatrix}\bigr)$ need not imply trivial couplings, as we will see.

Finally, consider the case $\vartheta^2 < 0$. Defining $\theta^2 \equiv -\vartheta^2 = -k_{11}^2 - k_{12} k_{21}$, one finds
\begin{equation}
  \Lambda_1 = \begin{pmatrix}
    \cos \theta + k_{11}  \frac{\sin \theta}{\theta} & k_{12}  \frac{\sin \theta}{\theta}\\
    k_{21}  \frac{\sin \theta}{\theta} & \cos \theta - k_{11}  \frac{\sin \theta}{\theta}
  \end{pmatrix} \in \SL (2, \mathbb{Z}) .
\end{equation}
Taking the trace, we conclude that $\cos \theta \in \{ 0, \pm
\frac{1}{2}, \pm 1 \}$, so that $\theta$ is a multiple of $\cpi / 2$
or $\cpi / 3$. For $\cos \theta \in \{ 0, \pm
\frac{1}{2}\}$, a general solution takes the form $k_{i j} = \frac{\theta}{2 \sin \theta} n_{i j}$ for integers $n_{ij}$ satisfying $n_{11}^2 + n_{12} n_{21} = 4 \cos^2 \theta - 4$ with $n_{12}$ and $n_{21}$ even. In other words,
\begin{equation}
  \begin{pmatrix}
    k_{11} & k_{12}\\
    k_{21} & - k_{11}
  \end{pmatrix} = \frac{\theta}{\sin \theta}  \begin{pmatrix}
    \frac{a_1 - d_1}{2} & b_1\\
    c_1 & \frac{d_1 - a_1}{2}
  \end{pmatrix}, \label{eqn:kLambdaEqn3}
\end{equation}
where $\theta$ is any solution to $\cos \theta = \frac{a_1 + d_1}{2}$. Thus, a
choice of monodromy matrix $\Lambda_1$ satisfying $\Tr \Lambda_1 \in \{
0, \pm 1 \}$ plus a choice of branch cut for $\theta = \cos^{- 1} \bigl(
\frac{\Tr \Lambda_1}{2} \bigr)$, $\theta > 0$ uniquely fixes the $k_{i
j}$.

Finally, in the case $\cos \theta = \pm 1$ there is no further constraint on the $k_{i j}$ beyond $k_{11}^2 + k_{12} k_{21} = -\theta^2$.

To summarize, the necessary and sufficient conditions for the $k_{i j}$
quantization rules to be satisfied are as follows. First, we pick a monodromy
matrix $\Lambda_1 \in \SL (2, \mathbb{Z})$ satisfying either $\Tr
\Lambda_1 \geqslant - 1$ or $\Lambda_1 = - 1_{2 \times 2}$. If $\Tr
\Lambda_1 \geqslant 2$ and $\Lambda_1 \neq 1_{2 \times 2}$ then this uniquely
fixes the $k_{i j}$ via \eqref{eqn:kLambdaEqn} or \eqref{eqn:kLambdaEqn2}. If
$- 1 \leqslant \Tr \Lambda_1 \leqslant 1$ then this fixes the $k_{i j}$
via \eqref{eqn:kLambdaEqn3} after a choice of branch cut for $\theta = \cos^{-
1} \bigl( \frac{\Tr \Lambda_1}{2} \bigr)$, $\theta > 0$. Finally, when
$\Lambda_1 = \pm 1_{2 \times 2}$ we require $k_{11}^2 + k_{12} k_{21} = - (n
\cpi)^2$ for $n \in \mathbb{Z}$ where $n$ is even (odd) when $\Lambda_1 = 1_{2
\times 2}$ ($\Lambda_1 = - 1_{2 \times 2}$). Note that only in the last case
can the $k_{i j}$ be varied continuously consistent with the quantization
rules. Otherwise they are discretely quantized.

\bibliography{ref}
\bibliographystyle{utphys}

\end{document}